\documentclass[%
 reprint,
 superscriptaddress,
 showkeys,
bibnotes,
 amsmath,amssymb,
 aps,
pra,
]{revtex4-1}
\usepackage{graphicx}
\usepackage{dcolumn}
\usepackage{bm}
\usepackage{siunitx}
\usepackage{booktabs}
\usepackage{makecell}
\usepackage[flushleft]{threeparttable}
\usepackage{natbib}
\usepackage{calligra}
\usepackage{subcaption}
\usepackage{amsmath}
\usepackage[dvipsnames]{xcolor}
\usepackage{mathrsfs}
\usepackage{xr}
\usepackage{hyperref}

\usepackage{listings}

\usepackage{color}
\definecolor{gray}{rgb}{0.4,0.4,0.4}
\definecolor{darkblue}{rgb}{0.0,0.0,0.6}
\definecolor{cyan}{rgb}{0.0,0.6,0.6}

\begin{document}


\title{Hydrodynamic signatures in thermal transport in devices based on 2D materials: an \textit{ab initio} study}


\author{Mart\'i Raya-Moreno}
\affiliation{Departament d'Enginyeria Electr\`onica, Universitat Aut\`onoma de Barcelona, 08193 Bellaterra, Barcelona,Spain}
\author{Jes\'us Carrete}
\affiliation{Institute of Materials Chemistry, TU Wien, A-1060 Vienna, Austria}
\author{Xavier Cartoix\`a}%
\email{Xavier.Cartoixa@uab.cat}
\affiliation{Departament d'Enginyeria Electr\`onica, Universitat Aut\`onoma de Barcelona, 08193 Bellaterra, Barcelona,Spain}%


\begin{abstract}
	We investigate the features arising from hydrodynamic effects in graphene and phosphorene devices with finite heat sources, using \textit{ab initio} calculations to go beyond Callaway's model and inform a full linearized scattering operator, and solving the phonon Boltzmann transport equation through energy-based deviational Monte Carlo methods. 
	We explain the mechanisms that create those hydrodynamic features, showing that boundary scattering and the relation of sample dimensions to the non-local length $\ell$ are the determinant factors, regardless of the relative importance of normal vs.\ resistive scattering. From this point of view, the non-local length $\ell$ reflects the ability
of scattering to randomize the heat flux, and we show that approximations made on the scattering operator may have, through the value of $\ell$, qualitative consequences on the signatures of hydrodynamic behavior.
\end{abstract}

\maketitle

\section{\label{sec:Intro}Introduction}

Over the last decades, the continuous development of micro- and nanofabrication techniques has allowed for higher integration levels in electronic devices and more efficient thermoelectric systems, as well as better thermal insulators for improved energy savings~\cite{Moores_law,MooreMatToday2014,MinnichEES2009,maldovanNature2013,TianSciAdv2018}. A proper understanding of the thermal transport in those systems becomes essential to optimize their operation since, if not carefully managed, heat can severely hinder their efficiency and/or durability~\cite{PopPIEEE2006,macii_book,panda_book,RossIEEESpec2008}.

Classically, heat transport has been modeled with Fourier's law~\cite{Fourier1822}. However, Fourier's law is known to break down in small systems and short timescales, with several examples available: for instance, it has been observed that Fourier's law fails to properly reproduce several thermoreflectance~\cite{WilsonNC2014,ZiabariNC2018,BeardoPRB2020} and thermal grating~\cite{JohnsonPRL2013,TorresPRM2018} experiments even with some, sometimes physically questionable, modifications~\cite{ZiabariNC2018}. Such erroneous descriptions have their origin in the shortcomings of Fourier's law, i.e. the infinite velocity of heat propagation~\cite{JosephRMP1989,ChristovPRL2005,PolThesis} and the lack of non-local effects~\cite{TorresPRM2018,PolThesis}. The inclusion of these two effects into the theory makes heat acquire characteristics of a viscous fluid, which are describable via additional hydrodynamic (Navier-Stokes-like) terms~\cite{GuyerPR1966,GuoPR2015,TorresPRM2018,PolThesis}.

In the case of semiconductors, where phonons are the main heat carriers, it is possible to obtain an accurate description of thermal transport, including non-Fourier features, by solving the Peierls-Boltzmann transport equation (PBTE) with the appropriate boundary conditions~\cite{ZimanEPH,LiPRB2012}. Indeed, if supplied with first-principles data, this latter equation can even model devices based on novel materials for which other simpler descriptions are lacking. Despite its advantages in terms of precision and transferability, the solution of the PBTE, for devices and/or structures of interest for heat management or thermoelectric applications, is rather complex and expensive. Moreover, the overall complexity of the PBTE makes it hard to obtain an intuitive and clear physical picture of heat transport phenomena~\cite{GuoPR2015,GuoPRB2018}.

In this context, beyond-Fourier mesoscopic models (i.e., the ones that not only describe the classical diffusive regime of Fourier's law, but other important transport regimes in the nanoscale, such as the ballistic and hydrodynamic ones) are essential for a fast, simple, and physically intuitive description of thermal transport at the micro/nanoscale~\cite{GuoPRB2018}. Among all available mesoscopic equations, the hydrodynamic equation stands out as a promising candidate since it can be directly derived from the BTE, so that a microscopic physical description of its variables might be obtained, enabling their calculation with \textit{ab initio} methods for accurate simulations~\cite{PolThesis,SendraPRB2021}.

Two clear signatures of hydrodynamic behavior are the formation of heat flux vortices and the appearance of non-monotonic temperature profiles in non-trivial sample geometries. Such hydrodynamic features have also been predicted to appear in the context of electronic transport for graphene~\cite{Levitov2016NP}, and some recent works have also predicted the existence of thermal vortices in nanoporous silicon~\cite{ZhangIJHMT2019} and graphene-based devices~\cite{Shang2020SR,ZhangIJHMT2021}, though the used methodology was, in these cases, based on fitted parameters or models instead of a full microscopic \textit{ab initio}-based description. This has been improved in the recent work of Guo \textit{et al.}~\cite{GuoIJHMT2021}, where they predicted vorticity in graphene-based devices based on Callaway's model~\cite{CallawayPR1959} informed with \textit{ab initio} scattering rates. However, despite the increase in the accuracy with respect to the gray model in Ref.~\cite{ZhangIJHMT2021} or the relaxation time approximation (RTA), Callaway's model is known to yield thermal conductivities with, at best, a $30\%$ error~\cite{DingPRB2018,GuoIJHMT2021} at room temperature in the case of graphene. Although the accuracy of Callaway's model can be improved by a more careful analysis of umklapp scattering in thermal resistivity~\cite{DingPRB2018,GuoPRB2021}, reducing the error to $3\%$, such improvement is not guaranteed \textit{a priori}. For instance, in the case of black phosphorous the error in the zigzag thermal conductivity, even with such a correction, is of $15\%$~\cite{DingPRB2018}. This lack of systematic predictability together with other theoretical drawbacks, like the requirement to differentiate between normal and umklapp processes---a distinction that some authors find arbitrary~\cite{taylor_heinonen_2002BOOK, MaznevAJP2014}---makes it necessary to go beyond and make use of the full linearized scattering operator for an accurate, predictive description of hydrodynamics in both real and reciprocal space. Although few scarce examples of such a step can be found in the literature~\cite{LandonJAP2014,LiPRB2018,LiPRB2019}, none of them discussed any complex geometries in which vorticity or non-monotonic temperature profiles appear. Consequently, this raises the question of to what extent simpler non-systematic approaches to the scattering operator, like Callaway's or the RTA, can properly describe such features, and how their use compares to the exact linearized collision operator. 

In this work, we study hydrodynamic signatures in graphene and phosphorene-based nanodevices at room temperature by solving the PBTE via energy-based deviational Monte Carlo (MC) techniques with a full linearized scattering operator where the scattering rates are obtained from first principles. We discuss the accuracy and effect on those hydrodynamic features of using non-systematic approaches to the scattering operator, namely the relaxation time approximation, in comparison to the most accurate full linearized operator. We also provide insight into the mechanisms originating those features, and we show that the key factor in vortex appearance is the vertical separation between the heat source and drain, regardless of the relative importance of normal vs.\ resistive scattering. Finally, we complement our results with solutions of the hydrodynamic equation for a better understanding of some size effects as well as studying signatures below the MC statistical noise.

The paper is structured as follows: after briefly discussing hydrodynamics in Sec.~\ref{sec:hydro2D}, we detail the applied methodology for the heat fluxes and thermal profiles calculation in Sec.~\ref{sec:Methods}, and we present the bulk properties, cumulative curves and a discussion of the expected hydrodynamic features in Sec.~\ref{sec:Bulk}. The results for devices are provided in Sec.~\ref{sec:Results}, both for shorter ballistic  (Sec.~\ref{sec:Results:Ballistic}) and taller devices (Sec.~\ref{sec:Results:NonBallistic}). Our summary and conclusions are given in Sec.~\ref{sec:Conclusions}.

\section{\label{sec:hydro2D}Hydrodynamics and 2D materials: the role of normal and umklapp processes}

The steady-state hydrodynamics equation of heat transport is given by:
\begin{equation}
\label{Eq:HYDRO}
{\mathbf{J}} - \ell^2 \nabla^2 \mathbf{J} = -\kappa \nabla T
\end{equation}
where $\mathbf{J}$ is the heat flux, $\kappa$ is the thermal conductivity tensor, $T$ is the temperature and $\ell$ is a non-local length tensor, which is related to the distance in which phonon distribution can conserve its inertia even under the effect of intrinsic scattering. 
One of the earliest derivations of this hydrodynamic equation for heat transport was by Guyer and Krumhansl (GK)~\cite{GuyerPR1966}. In their work, they derived it by expanding the phonon distribution in eigenvectors of the normal collision operator under the condition that normal ($\mathscr{N}$) scattering is dominant over umklapp ($\mathscr{U}$) and extrinsic ($\mathscr{E}$) scattering. Indeed, within the GK framework, hydrodynamic features are only possible when $\mathscr{N}$ processes dominate over all other kinds of scattering~\cite{GuoPR2015,CepellottiNatCom2015, Shang2020SR, ChenNatRP2021}. 

A common approach to the scattering operator when solving the PBTE is the so-called relaxation time approximation (RTA), where it is assumed that each phonon mode relaxes to equilibrium independently. Despite its simplicity, such an approach has been observed to properly describe thermal properties for prototypical semiconductors like silicon. However, since in the RTA all scattering processes are deemed resistive, the RTA is expected to fail for cases in which most of the processes are not directly resistive but indirectly so through population redistribution, which has been usually associated to $\mathrm{\mathscr{N}~\gg~\mathscr{U},\mathscr{E}}$~\cite{GuyerPR1966,CepellottiNatCom2015}. Thus, in 2D materials, in which the RTA has been observed to provide a poor description of thermal properties~\cite{CepellottiNatCom2015,LindsayJAP2019}, hydrodynamic features should be stronger than for other materials~\cite{CepellottiNatCom2015,Shang2020SR}.

It is worth noting that the role of $\mathscr{U}$ and $\mathscr{N}$ processes in phonon hydrodynamics is more nuanced than the widespread belief, rooted in the GK derivation, that $\mathscr{N}$ processes must dominate~\cite{CepellottiNatCom2015, Shang2020SR, ChenNatRP2021}. That is a sufficient but not necessary condition, since the hydrodynamic regime---which we will associate to non-ballistic systems where the $\ell^2 \nabla^2 \mathbf{J}$ term produces appreciable deviations from Fourier's law---is possible even in the case where $\mathscr{N}$ and $\mathscr{U}$ scattering are comparable or when $\mathscr{E}$ dominates over intrinsic scattering~\cite{GuoPR2015, ZhangIJHMT2021, SendraPRB2021}. In fact, by projecting the PBTE over energy and quasimomenta, and by expanding the phonon distribution on macroscopic variables like the flux, Sendra~\textit{et al.}~\cite{SendraPRB2021} have derived a general hydrodynamic formalism that does not rely on the $\mathscr{N}$ dominance or on the classification of three-phonon processes into $\mathscr{N}$ and $\mathscr{U}$, which is subject to the specific choice of the primitive cell~\cite{taylor_heinonen_2002BOOK, MaznevAJP2014}. This generalizes hydrodynamics to materials where both resistive and non-resistive processes are comparable---i.e.: out of the range of applicability of the RTA---, as in the case of diamond or 2D materials~\cite{GuoPR2015}, but also to materials like room-temperature silicon in which intrinsic resistive scattering is the dominant mechanism~\cite{SendraPRB2021}. Then, the manifestation of the hydrodynamic regime will depend on the value of $\ell$ for a specific material and experimental configuration.

Taking into consideration everything mentioned here, we have selected graphene and phosphorene for our study. The former was chosen because, in addition to it being the prototypical 2D material, several works~\cite{Levitov2016NP,Shang2020SR} have already predicted current vortices in both electronic and thermal transport. On the other hand, the latter was chosen in view of its excellent electronic properties~\cite{HaratipourACSNano2016}, which have positioned it as a principal actor in the survival of Moore's law down to atomic sizes~\cite{IlatikhamenehSR2016}, thus making a proper description of heat transport, including hydrodynamics, essential for the developing of phosphorene-based devices.

\section{\label{sec:Methods}Methodology}

\begin{figure}[ht!]
	\centering
	\includegraphics[width=\linewidth]{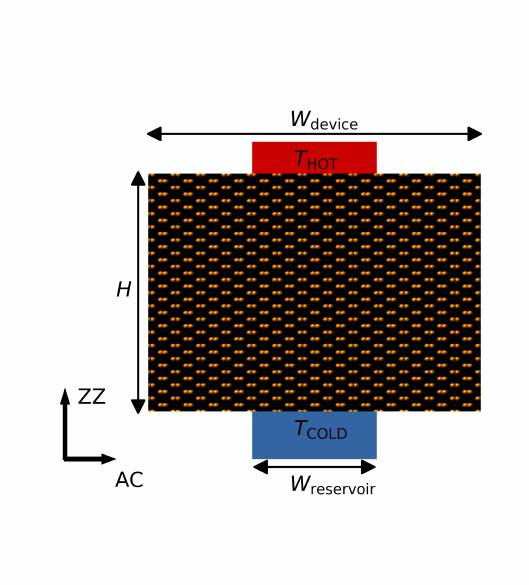}
	\caption{Sketch of a Levitov configuration with characteristic lengths $H$, $W_{\mathrm{reservoir}}$ and $W_{\mathrm{device}}$ indicated. The transport axis, armchair (AC) and zigzag (ZZ), for phosphorene case are given as reference.}
	\label{LEVISKETCH}
\end{figure}

Previous works have predicted electrical current vortices and negative nonlocal resistance in 2D materials (graphene) when finite sources are injecting heat/electrons~\cite{Levitov2016NP,Shang2020SR,ZhangIJHMT2021}. A sketch of those geometries, henceforth called Levitov configurations, is depicted in Fig.~\ref{LEVISKETCH}.
The dimensions of the studied Levitov configurations were selected to investigate all possible regimes; to that end, a mean free path ($\lambda_\alpha$) and a non-local length ($\ell$) have been computed to provide an approximate idea of the limiting sizes of each transport regime. The mean free path is defined as
\begin{equation}
\lambda_\alpha \equiv ([\kappa^{\mathrm{SG}}]^{-1}\kappa^{\mathrm{RTA}})_{\alpha\alpha}.
\end{equation}

Here $\alpha$ indicates a Cartesian axis, $\kappa^{\mathrm{RTA}}$ is the RTA thermal conductivity tensor and $\kappa^{\mathrm{SG}}$ is the small-grain thermal conductivity defined as $\kappa^{\mathrm{SG}} = \frac{1}{k_BT^2V_{uc}}\sum_\sigma n^0_\sigma(n^0_\sigma + 1)\frac{\hbar\omega_\sigma}{|v_\sigma|}v_\sigma \otimes v_\sigma $~\cite{ShengBTE}, where $k_B$ is the Boltzmann constant, $V_{uc}$ is the volume of the unit cell, $\sigma$ is a phonon mode, $n^0_\sigma$ is the Bose-Einstein distribution for mode $\sigma$, $\hbar$ is the reduced Planck constant, $\omega_\sigma$ is the phonon frequency and $v_\sigma$ is the phonon group velocity. In addition, the non-local length ($\ell$) within the RTA and for isotropic materials, $\ell^\mathrm{RTA}_\mathrm{iso}$, can be calculated as~\cite{SendraPRB2021}:
\begin{equation}
({\ell^\mathrm{RTA}_\mathrm{iso}})^2 = \frac{1}{5}\frac{\int_{BZ} \hbar q_\sigma v_\sigma^2 \tau_\sigma^2 \partial_T n^0_\sigma d\sigma}{\int_{BZ} \hbar q_\sigma v_\sigma \partial_T n^0_\sigma d\sigma},
\end{equation}
where $q_\sigma$ is the phonon quasimomentum, $\tau_\sigma$ is the phonon lifetime and $\partial_T n^0_\sigma$ is the derivative with respect to temperature of the Bose-Einstein distribution. 
Additionally, we also provide the thermal conductivity as it is an essential quantity in the hydrodynamic equation. Cumulative quantities were also computed as a mean to obtain a deeper understanding of size effects in these configurations~\cite{LiPRB2012}.

Heat flux and temperature profiles for several Levitov configurations are obtained through the PBTE solution. Owing to the high complexity of the Levitov geometry, we use an efficient, energy-based deviational Monte Carlo approach~\cite{PeraudPRB2011,PeraudAPL2012,PeraudARHT2014,LandonJAP2014,LandonThesis} to solve the PBTE under the RTA and beyond (bRTA), as implemented in \verb|BTE-Barna|~\cite{myself} to solve the PBTE.

Atomic positions, harmonic and anharmonic interatomic force constants (IFCs), needed to compute the basic phonon properties (i.e.: group velocities, lifetimes, frequencies) and the propagator~\cite{LandonJAP2014,myself} required for the PBTE Monte Carlo solvers, were obtained from Refs.~\cite{almadatabase} for graphene and~\cite{SmithAM2017} for phosphorene. Harmonic IFCs were corrected to enforce crystal symmetry, translational invariance, and rotational invariance, necessary for a proper description of quadratic acoustic bands~\cite{CarreteMRL2016}, the broadening parameter was fixed to $1$ for energy conservation and the phosphorene and graphene layer thicknesses were set to \SI{0.533}{\nano\meter}~\cite{CastellanosGomez2DMat2014} and \SI{0.335}{\nano\meter}~\cite{LandonJAP2014} respectively. The phosphorene (graphene) phonon properties and the propagator were calculated on a $\Gamma$-centered $q$-mesh of $50\times50\times1$ ($80\times80\times1$) points, for which the thermal bulk conductivity is found to be converged---within 5\% with respect to a higher quality mesh of $100\times100\times1$ ($110\times110\times1$) points---at \SI{300}{\kelvin}. For both materials, the propagator was calculated using a time step of \SI{0.25}{\pico\second} and a reference temperature of \SI{300}{\kelvin}. 

Finally, to efficiently investigate some dimensional limits and other features as the existence of vortices with magnitudes well below statistical noise caused by the intrinsic scattering algorithms~\cite{PeraudARHT2014,myself}, we solved Eq.~(\ref{Eq:HYDRO}) using finite elements package FEniCS~\cite{FeniCsBook,alnaes2015fenics}, with its parameters extracted from cumulative curves (see the \verb|fenics.py| script in the Supplemental Material~\cite{SM} for further details on those calculations).

\section{\label{sec:Bulk}Bulk properties and size effects}

In their work, Shang \textit{et al.}~\cite{Shang2020SR} proposed two different mechanisms as the origin of thermal vortices. For small systems, in the ballistic regime, the source would be the combination of elastic boundary scattering plus the phase of phonons but not hydrodynamics per se. On the other hand,  for larger devices, they concluded that vorticity is due to $\mathscr{N}$-scattering dominance together with boundary conditions (finite sources injecting the heat into the system).

\begin{figure*}[ht!]
	\begin{subfigure}[b]{0.45\textwidth}
		\centering
		\includegraphics[width=\textwidth]{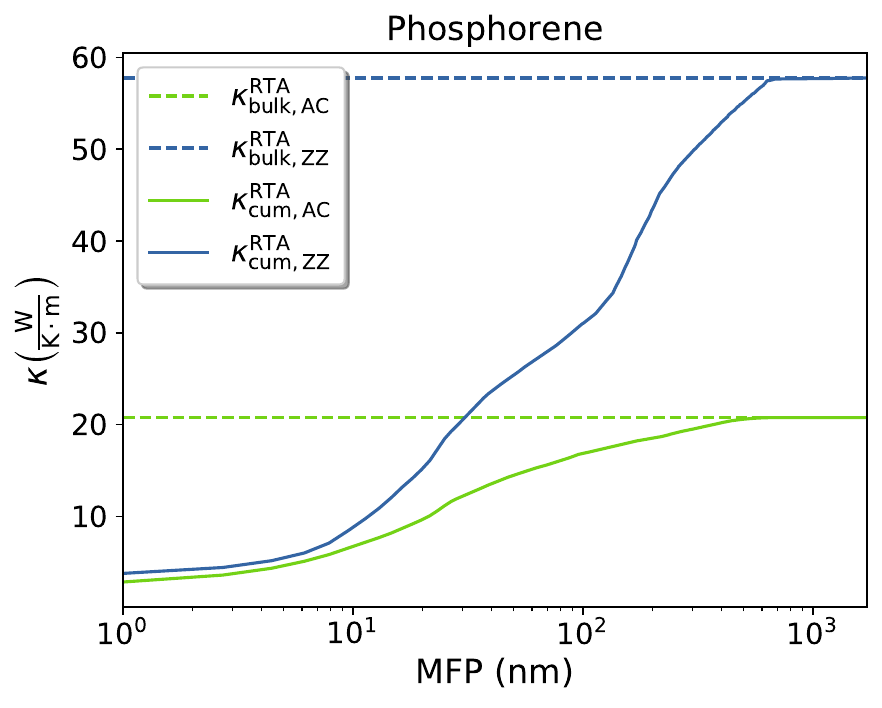}
		\label{SFig:KAPPACUM_bP}
	\end{subfigure}
	\begin{subfigure}[b]{0.45\textwidth}
		\centering
		\includegraphics[width=\textwidth]{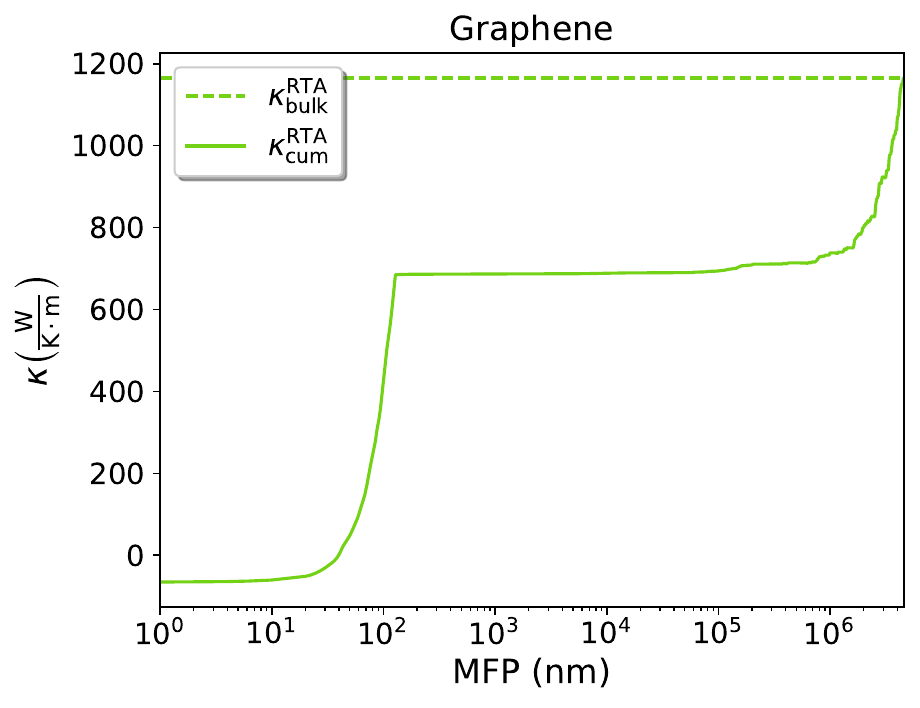}
		\label{SFig:KAPPACUM_gph}
	\end{subfigure}
	\caption{Cumulative $\kappa^\mathrm{RTA}$ for phosphorene (left) and graphene (right) with respect to mean free path at \SI{300}{\kelvin}. Bulk values are provided as reference. These values have been computed in a denser $\mathbf{q}$-mesh, namely $300\times300\times1$ ($480\times480\times1$) for phospherene (graphene), through a cubic spline interpolation from the finer mesh to remove nonphysical artifacts.}
	\label{Fig:KAPPACUM}
\end{figure*}

\begin{figure*}[ht!]
	\begin{subfigure}[b]{0.45\textwidth}
		\centering
		\includegraphics[width=\textwidth]{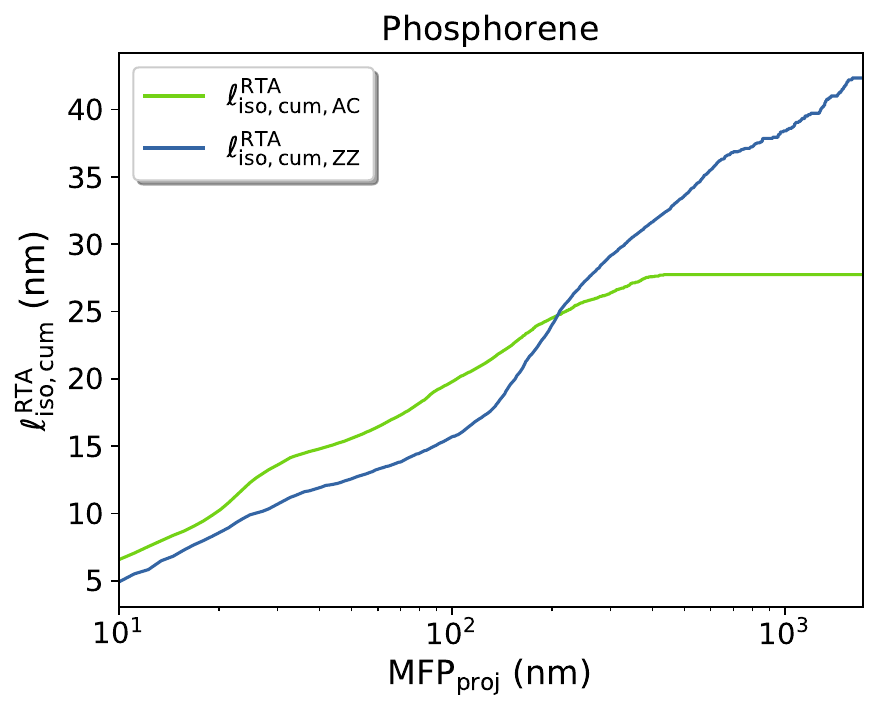}
		\label{SFig:LCUM_bP}
	\end{subfigure}
	\begin{subfigure}[b]{0.45\textwidth}
		\centering
		\includegraphics[width=\textwidth]{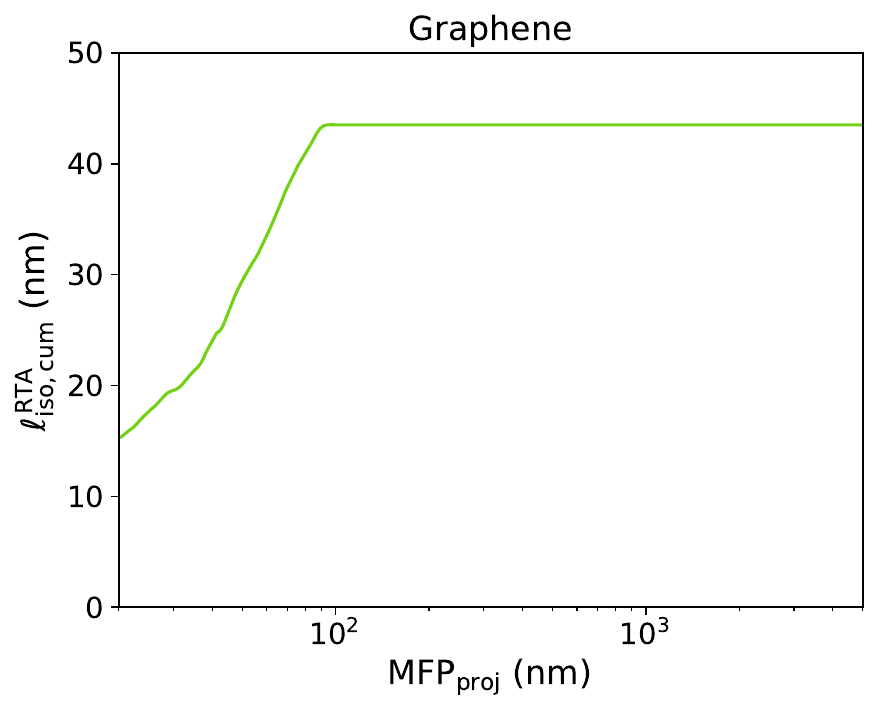}
		\label{SFig:LCUM_gph}
	\end{subfigure}
	\caption{Cumulative $\ell^\mathrm{RTA}_\mathrm{iso}$ for phosphorene (left) and graphene (right) with respect to projected mean free path at \SI{300}{\kelvin}. The graphene cumulative function has been restricted to small $\lambda$ as almost all the contribution to bulk values comes from modes with $\lambda$ around \SI{4.2}{\milli\meter}. These values have been computed in a denser $\mathbf{q}$-mesh, namely $300\times300\times1$ ($480\times480\times1$) for phospherene (graphene), through a cubic spline interpolation from the finer mesh to remove nonphysical artifacts.}
	\label{Fig:LCUM}
\end{figure*}

To get an idea of the interplay between the device dimensions and its transport regime, we have calculated the bulk non-local length and mean free path for graphene and phosphorene at room temperature (see Table~\ref{Tab:BULK}). Additionally, we also provide the thermal conductivity because of its importance for heat hydrodynamics, with very good agreement with other theoretical calculations and experimental values.
Although the non-local lengths and mean free paths offer a general idea of transport regime limits, they fail to provide insight on how system boundaries partially suppress the macroscopic thermal properties (i.e. $\kappa$ and $\ell$), as longer-mean-free-path phonons do not fully contribute to these macroscopic variables~\cite{ZimanEPH}. The last is especially important as it modifies $\ell$, which is one of the quantities determining the transport regime.
While it would be possible to obtain the effective macroscopic quantities for given geometries via averaging~\cite{LiPRB2012}, the overall complexity of the Levitov configuration makes such a task unfeasible. On the other hand, even though, as noted by Li \textit{et al.}~\cite{LiPRB2012}, mean free path cumulative functions do not hold accurate predictive power, they are a useful tool to interpret size effects on transport variables. To that end, we provide the cumulative thermal conductivity and non-local length as a function of phonon mean free path (see Figs.~\ref{Fig:KAPPACUM}~and~\ref{Fig:LCUM}).

\begin{table}
	\renewcommand{\arraystretch}{0.23} 
	\centering
	\caption{\label{Tab:BULK} Calculated $\kappa$, $\ell$ and $\lambda$ for bulk graphene and bulk phosphorene at \SI{300}{\kelvin}. The experimental values of $\kappa$ and values of other theoretical calculations are also provided.}
	\begin{threeparttable}
		\begin{tabular}{@{}llllll@{}}
			\toprule
			\hline \hline \\
			& \makecell[tc]{$\kappa^\mathrm{RTA}$ \\ $\mathrm{\left(\frac{W}{m\cdot K}\right)}$} & \makecell[tc]{$\kappa^\mathrm{bRTA}$ \\ $\mathrm{\left(\frac{W}{m\cdot K}\right)}$} &  \makecell[tc]{$\kappa^\mathrm{ref}$ \\ $\mathrm{\left(\frac{W}{m\cdot K}\right)}$} & \makecell[tc]{$\ell^\mathrm{RTA}_\mathrm{iso}$ \\ $\left(\mathrm{nm}\right)$} &
			\makecell[tc]{$\lambda{\;}$ \\ $\left(\mathrm{nm}\right)$}\\ \\  \midrule
			\hline \\
			\\
			\makecell[tl]{Graphene} & \makecell[tc]{ 1.17\\$\times10^3$} & \makecell[tc]{4.20\\$\times10^3$} & \makecell[tc]{(4.84-5.30)\\ $\times10^3$$\,^a$ \\ (3.08-5.15)\\$\times10^3$$\,^{b}$} &\makecell[tc]{5.39\\$\times10^4$\\(43.5$^\dagger$)} & \makecell[tc]{169} \\
			\\
			\makecell[tl]{Phosphorene \\ (AC)} & \makecell[tc]{20.8} & \makecell[tc]{27.5} & \makecell[tc]{23.9$^{c}$ \\ 35.5$^{d}$ \\ 22.0$^{e}$} &\makecell[tc]{27.7} & \makecell[tc]{20.2} \\
			\\
			\makecell[tl]{Phosphorene \\ (ZZ)} & \makecell[tc]{57.8} & \makecell[tc]{82.9} & \makecell[tc]{82.1$^{c}$ \\ 108$^{d}$ \\ 63.2$^{e}$} &\makecell[tc]{42.3} & \makecell[tc]{33.5} \\
			\\
			&  &  &  \\ \bottomrule
			\hline \hline
		\end{tabular}
		\begin{tablenotes}
			\small
			\item  $^\dagger$ this value corresponds to the $\ell^\mathrm{RTA}_\mathrm{iso}$ calculation without including low frequency ZA modes in the neighborhood of $\Gamma$ with $\lambda$ around \SI{4.2}{\milli\meter}.
			\item $^a$ Experimental $\kappa$ obtained from Ref.~\cite{BalandinNL2008}.
			\item $^b$ Experimental $\kappa$ obtained from Ref.~\cite{GhoshAPL2008}.
			\item $^c$ First-principles calculated $\kappa$ obtained from Ref.~\cite{ZhuPRB2014}. $\kappa$ is rescaled to take into account differences in assumed thickness.
			\item $^d$ First-principles calculated $\kappa$ obtained from Ref.~\cite{JainSR2015}. $\kappa$ is rescaled to take into account differences in assumed thickness.
			\item $^e$ First-principles calculated $\kappa$ obtained from Ref.~\cite{SmithAM2017}.
		\end{tablenotes}
	\end{threeparttable}
\end{table}

It should be noted that the theoretical formulas used to obtain the $\ell^\mathrm{RTA}_\mathrm{iso}$ values are derived under the assumption of isotropy and are therefore expected to be useful only as approximations in the phosphorene case. Regarding $\ell$ values beyond the RTA, we expect them to be higher overall, as in the RTA all processes are incorrectly deemed as directly resistive~\cite{SendraPRB2021,myself}. Notwithstanding this, in the case of graphene the dominance over $\ell^\mathrm{RTA}_\mathrm{iso}$ of low-frequency ZA-modes in the neighborhood of $\Gamma$ with extremely large mean free paths (of the order of \SI{4}{\milli\meter})---to the point that when suppressed the $\ell^\mathrm{RTA}_\mathrm{iso}$ gets reduced from \SI{53.9}{\micro\meter} to \SI{43.5}{\nano\meter}---makes the qualitative prediction of the beyond-RTA-$\ell$ behavior much more complicated, as the proper description of scattering operator would highly affect those ZA-modes. This is not surprising, as the RTA is known to fail in predicting graphene's thermal properties~\cite{CepellottiNatCom2015}. For instance, in our case the RTA predicts $\kappa$ to be only $\mathrm{27\%}$ of the fully converged value.

Regarding device dimensions, in the phosphorene case we expect hydrodynamic features to be observable out of the (quasi)ballistic regime, as the mean free path is a bit longer than the non-local length, the true value of which is expected to be larger than the one in Table~\ref{Tab:BULK}~\cite{myself}.  The latter suggests that phosphorene is a highly hydrodynamic material at the nanoscale, as the phonon distribution is capable of keeping its inertia even under the effect of significant intrinsic scattering. 
The case of graphene is somewhat more complex, as the major contribution to thermal variables, especially the non-local length, comes from the low-frequency ZA modes in the neighborhood of $\Gamma$ with extremely large mean free paths. At the nanoscale, the contribution of these modes to thermal conductivity or non-local length becomes negligible because a large suppression factor due to boundary scattering---see Eq.~(14) in Ref.~\cite{ShengBTE}---, thus lowering the non-local length to \SI{43.5}{\nano\meter}. Hence for graphene nanodevices, since that non-local length is smaller than $\lambda$, one would only expect hydrodynamics within the (quasi)ballistic regime.

\section{\label{sec:Results}Results}

Before presenting the results of our simulations, we find it worth commenting on the hydrodynamic signatures we have analyzed. Although the hydrodynamic regime presents several characteristic features, based on previous results~\cite{Levitov2016NP,Shang2020SR,ZhangIJHMT2021} we have focused our study on flux vortices and negative thermal resistance regions, namely regions in which the flux propagates in the same direction as the thermal gradient. None of those features can be easily described using Fourier's law unless one relies on complex models allowing for a position-dependent non-diagonal and non-definite-positive $\kappa$-tensor. However, these features arise naturally in hydrodynamics (see Eq.~\ref{Eq:HYDRO}) without using those complex and sometimes unphysical models, just unique and diagonal $\kappa$ and $\ell$ tensors.

\begin{figure}[ht!]
	\centering
	\includegraphics[width=\linewidth]{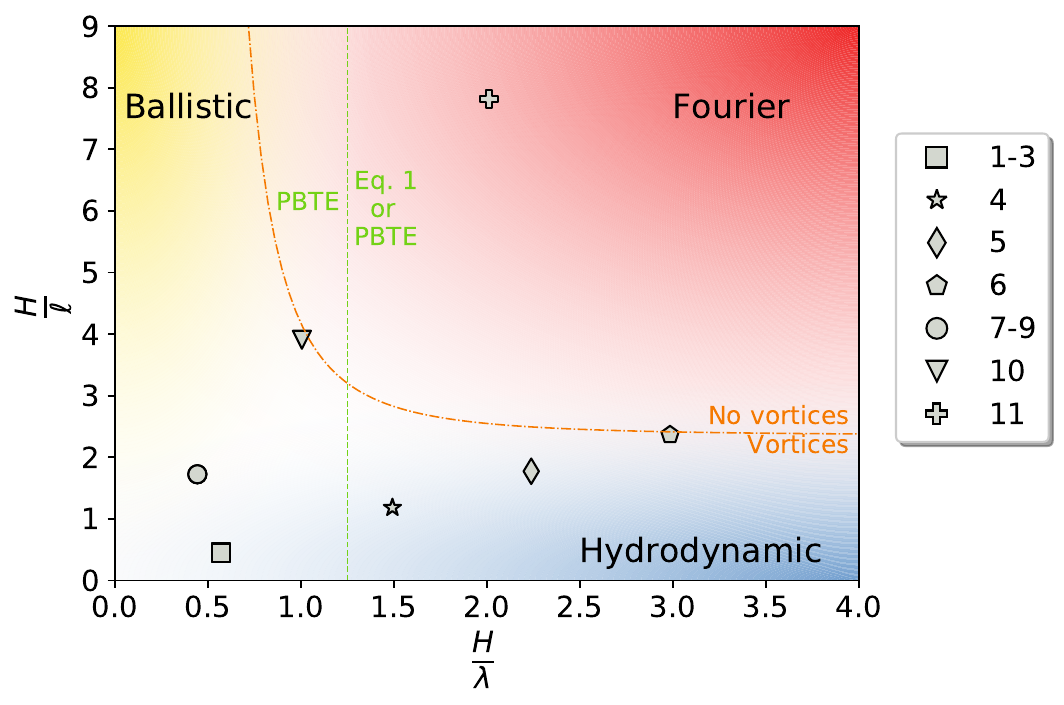}
	\caption{Ballistic and hydrodynamic regime scores at \SI{300}{\kelvin} for all the different Levitov structures studied in this work (see Table~\ref{Tab:Sizes} for the id. reference). The transport regime for the given scores is indicated with the background color---i.e. Fourier (red), hydrodynamic (blue) or ballistic (yellow)---with gradient zones representing a transition between the different regimes. The green dashed line gives an idea of the crossover between ballistic and regimes for which Eq.~\ref{Eq:HYDRO} can provide a more quantitative prediction; namely regions without strong ballistic features which are describable through the PTBE but also the hydrodynamic equation. The orange dashed line provides a approximate idea of for which scores we expect vortices to be possible for acceptable values of pipe-flow score (i.e. $<0.5$).}
	\label{Fig:Regimes}
\end{figure}

Furthermore, we also find it worth providing a summary of the characteristics of all the devices presented in this Section with an indication of the transport regime they are expected to be found on (see Fig.~\ref{Fig:Regimes} and Table~\ref{Tab:Sizes}). In the next Subsections, we discuss the transition from ballistic to non-ballistic regimes together with hydrodynamic to diffusive or Fourier regimes and their crossovers. Additionally, we provide insight on the effect of the source sizes with respect to device width. Finally, we use the graphene wider devices to obtain information regarding the formation of the hydrodynamic features.  

\begin{table*}[ht!]
	\centering
	\caption{\label{Tab:Sizes} Configuration id., material, characteristic sizes and figure references to all the different Levitov structures studied in this work. To illustrate the expected regimes of each device we provide ballistic and hydrodynamic regime scores at \SI{300}{\kelvin}, with lower values representing to be well inside the respective regime. Boldface in ballistic and hydrodynamic scores indicates  that, within the RTA, such device is well inside in that respective transport regime. Additionally, we provide a pipe-flow score, meaning 0 point sources and 1 a pipe like structure, so that for higher values vorticity is not allowed as in this limit one recovers a Poiseuille-like flow.}
	\begin{threeparttable}
		\begin{tabular}{@{}cccccccccc@{}}
			\toprule
			\hline \hline \\
			& Id. \#
			& Material  
			&\makecell[tc]{$W_{\mathrm{reservoir}}$ \\ $\mathrm{\left(nm\right)}$}
			&\makecell[tc]{$W_{\mathrm{device}}$ \\ $\mathrm{\left(nm\right)}$}
			&\makecell[tc]{$H$ \\ $\mathrm{\left(nm\right)}$}
			&\makecell[tc]{Ballistic \\ i.e. ${\frac{H}{\lambda}}^\dagger$}
			&\makecell[tc]{Hydrodynamic \\ i.e. ${\frac{H}{\ell}}^\dagger$}
			&\makecell[tc]{Pipe-flow \\ i.e. $\frac{W_{\mathrm{reservoir}}}{W_{\mathrm{device}}}$}
			&Figs.
			\\ \\  \midrule
			\hline \\
			\\
			& 1 & Phosphorene & 25.0 & 160           & 19.2 & \textbf{0.573}    & \textbf{0.454}  & 0.156 & \ref{Fig:PbalDev},\ref{Fig:VORTEXPHOSPHORENE:small} \\
			& 2 & Phosphorene & 25.0 & 320           & 19.2 & \textbf{0.573}    & \textbf{0.454}  & 0.078 & \ref{Fig:VORTEXPHOSPHORENE:large} \\
			& 3 & Phosphorene & 25.0 & 46.0          & 19.2 & \textbf{0.573}    & \textbf{0.454}  & 0.543 & \ref{Fig:limit} \\
			& 4 & Phosphorene & 25.0 & 160           & 50.0 & 1.492    & \textbf{1.182}  & 0.156 & \ref{Fig:P50},\ref{Fig:T50} \\
			& 5 & Phosphorene & 25.0 & 160           & 75.0 & 2.239    & \textbf{1.773}  & 0.156 & \ref{Fig:P75},\ref{Fig:T75} \\
			& 6 & Phosphorene & 25.0 & 160           & 100  & 2.985    & 2.364  & 0.156 & \ref{Fig:P100},\ref{Fig:T100} \\
			& 7 & Graphene    & 37.5 & 375           & 75.0 & \textbf{0.444}    & \textbf{1.724}  & 0.100 & \ref{Fig:GPHbalDev},\ref{Fig:VortexFormation} \\
			& 8 & Graphene    & 37.5 & 750           & 75.0 & \textbf{0.444}    & \textbf{1.724}  & 0.050 & \ref{Fig:DoubleVortexFD3} \\
			& 9 & Graphene    & 37.5 & 3750 & 75.0 & \textbf{0.444}    & \textbf{1.724}  & 0.010 & \ref{Fig:VortexFormationLARGE} \\
			& 10 & Graphene    & 37.5 & 375           & 170  & 1.006    & 3.908  & 0.100 & \ref{Fig:GPHnoBAL} \\
			& 11 & Graphene    & 37.5 & 375           & 340  & 2.012    & 7.816  & 0.100 & \ref{Fig:GPHnoBALLARGE} \\
			&  &  &  \\ \bottomrule
			\hline \hline
		\end{tabular}
		\begin{tablenotes}
			\small
			\item  $^\dagger$ For phosphorene devices the scores are calculated using the $\mathrm{ZZ}$ values of $\ell$ and $\lambda$, as it is the leading direction of thermal transport.
		\end{tablenotes}
	\end{threeparttable}
\end{table*}

\subsection{\label{sec:Results:Ballistic}Quasiballistic devices}

\begin{figure*}[ht!]
	\centering
	\includegraphics[width=0.9\textwidth]{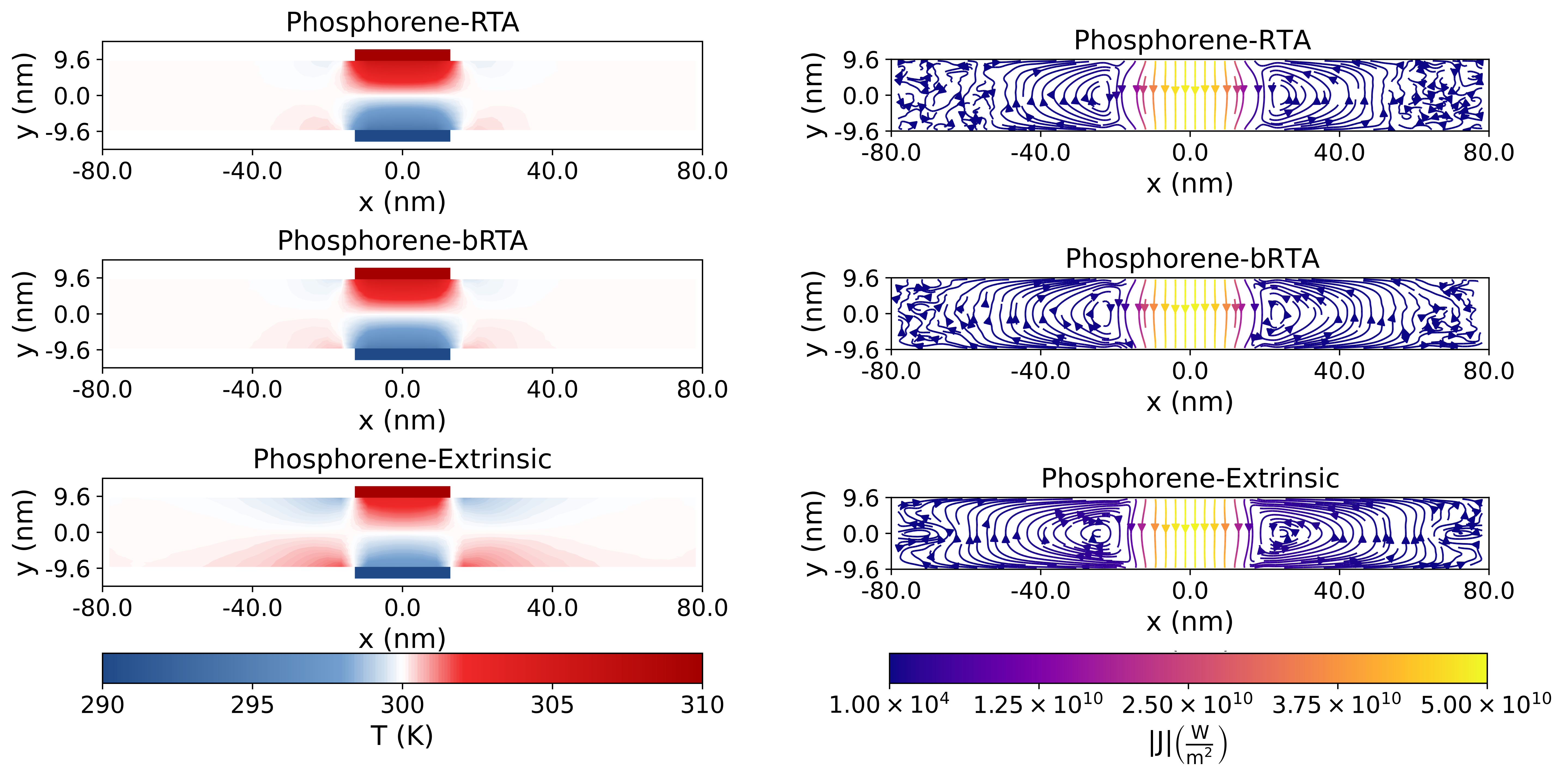}
	\caption{The RTA (top), bRTA (middle) and extrinsic (bottom) steady-state thermal profiles (left) and heat fluxes (right) for a ballistic phosphorene-based Levitov configuration with $W_{\mathrm{reservoir}}=$ \SI{25}{\nano\meter}, $W_{\mathrm{device}}=$ \SI{160}{\nano\meter} and $H=$ \SI{19.2}{\nano\meter}, $T_{\mathrm{HOT}}=$ \SI{310}{\kelvin} and $T_{\mathrm{COLD}}=$ \SI{290}{\kelvin}. The dark-scarlet red and dark-sky blue boxes at the top and bottom in the temperature profile panels are the hot and cold isothermal reservoirs, respectively.}
	\label{Fig:PbalDev}
\end{figure*}

\begin{figure*}[ht!]
	\centering
	\includegraphics[width=0.9\textwidth]{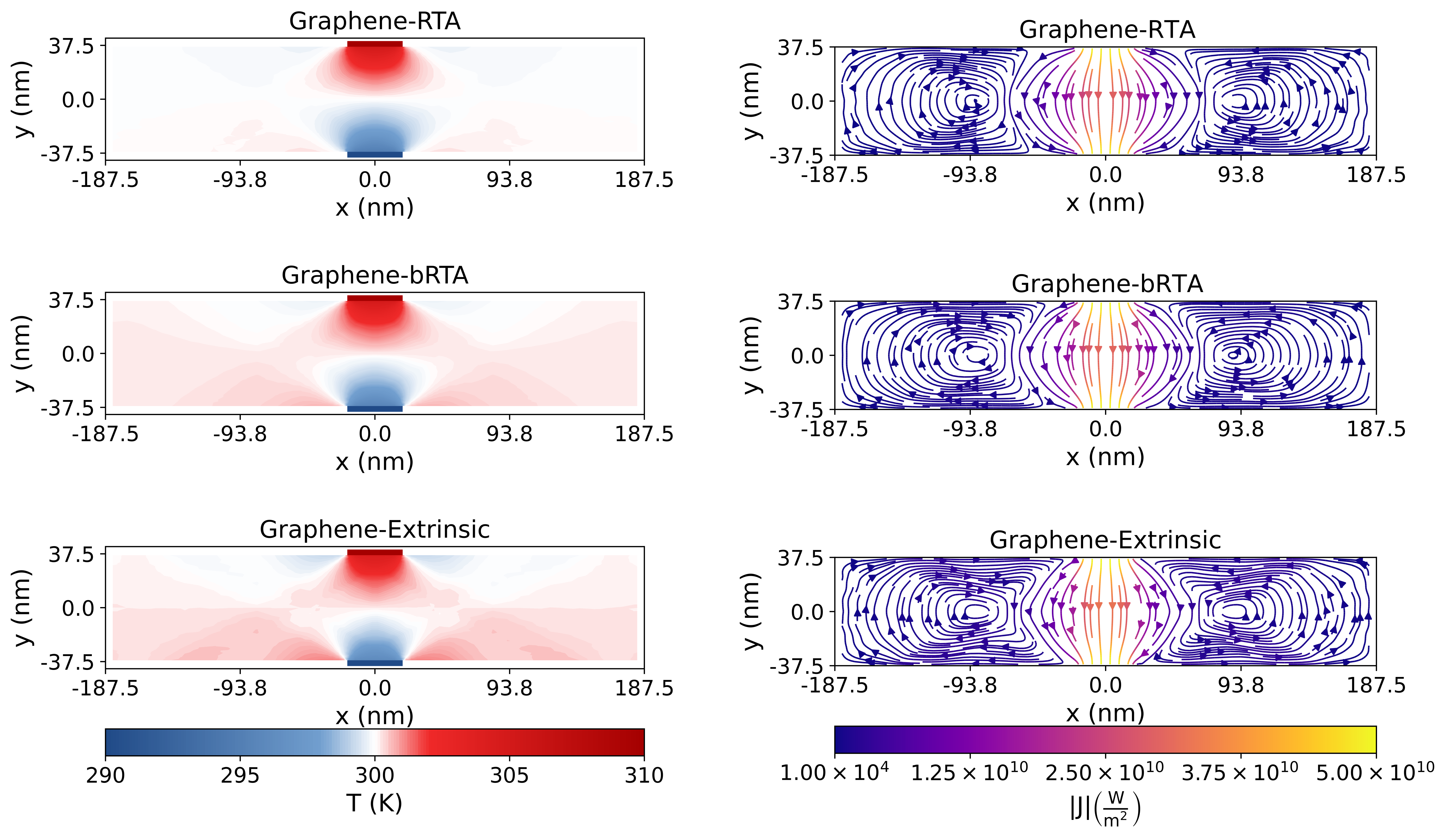}
	\caption{The RTA (top), bRTA (middle) and extrinsic (bottom) steady-state thermal profiles (left) and heat fluxes (right) for a ballistic graphene-based Levitov configuration with $W_{\mathrm{reservoir}}=$ \SI{37.5}{\nano\meter}, $W_{\mathrm{device}}=$ \SI{375}{\nano\meter} and $H=$ \SI{75}{\nano\meter}, $T_{\mathrm{HOT}}=$ \SI{310}{\kelvin} and $T_{\mathrm{COLD}}=$ \SI{290}{\kelvin}. The dark-scarlet red and dark-sky blue boxes at the top and bottom in the temperature profile panels are the hot and cold isothermal reservoirs, respectively.}
	\label{Fig:GPHbalDev}
\end{figure*}

Here, we provide the thermal profile~\footnote{In this and all the subsequent plots the color scale of the temperature profiles is sightly non-linear in the vicinity of 300 K to highlight the negative resistivity zones} and heat flux for phosphorene and graphene-based Levitov configurations well inside quasiballistic regime (see Figs.~\ref{Fig:PbalDev} and~\ref{Fig:GPHbalDev}), i.e. $H < \lambda$, for within and beyond the RTA. 

All temperature profiles show a small linear variation of temperature between the reservoirs (i.e. $x=0$), with abrupt changes near the reservoirs. Such a feature confirms that our devices are in the quasiballistic regime~\cite{MeiJAP2014}. This statement is further supported by heat flux profiles, in which the small differences in fluxes between the bRTA and RTA indicate the dominance of boundary scattering (as this mechanism is not dependent on the approach used to describe intrinsic scattering) thus devices are in the quasiballistic regime~\cite{CepellottiNatCom2015,ChenNatRP2021}. 

Regarding the hydrodynamic signatures, we can observe two flux vortices for all devices. Moreover, by combining flux and thermal profiles, we observe zones of negative thermal resistance at both sides of the reservoirs (see hot (cold) regions at both sides of cold (hot) reservoirs with the heat flux going downwards in Figs.~\ref{Fig:PbalDev} and \ref{Fig:GPHbalDev}). Thus, all our (quasi)ballistic devices are in the hydrodynamic regime, even in the absence of $\mathscr{N}$ processes, in opposition to more classical hydrodynamic theories.

Finally, the observable differences between graphene and phosphorene, especially for heat fluxes, which are as expected of higher values in graphene, are easily understood by noting the higher group velocities and lower scattering rates of graphene when compared to the phosphorene ones. Moreover, the asymmetric vortices found in phosphorene have their root in the material anisotropy.

\begin{figure}[ht!]
	\centering
	\includegraphics[width=0.45\textwidth]{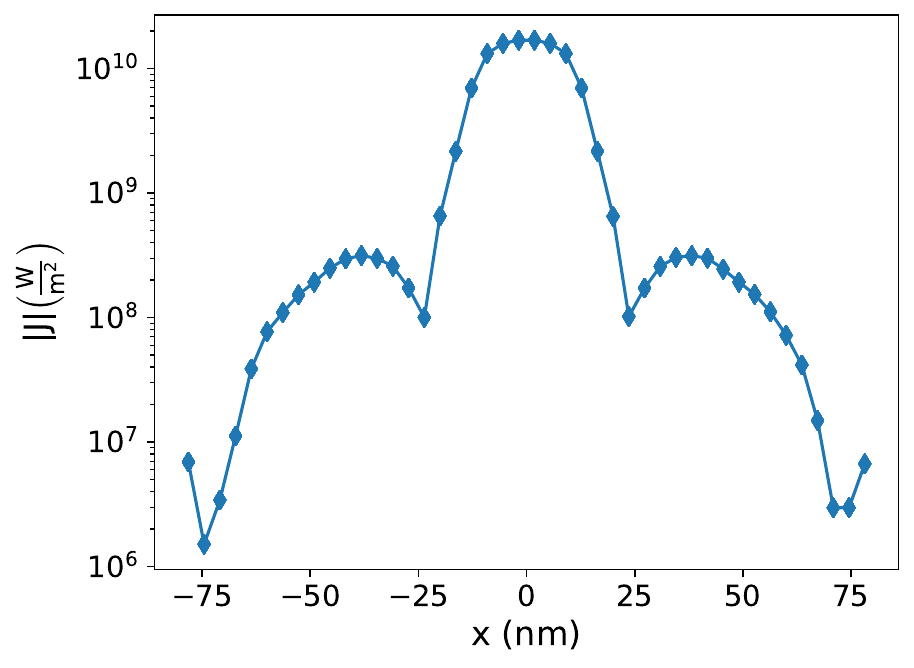}
	\caption{Cut of heat flux magnitude at extrinsic case of Fig.~\ref{Fig:PbalDev} for $y$=\SI{0}{\nano\meter} as function of $x$.}
	\label{Fig:DoubleVortexbP}
\end{figure}

\begin{figure*}[ht!]
	\centering
	\begin{subfigure}[b]{0.485\textwidth}
		\captionsetup{labelformat=empty}
		\caption{}
		\label{Fig:VORTEXPHOSPHORENE:small}
	\end{subfigure}
	\begin{subfigure}[b]{\textwidth}
		\captionsetup{labelformat=empty}
		\caption{}
		\label{Fig:VORTEXPHOSPHORENE:large}
	\end{subfigure}
	\begin{subfigure}[b]{\textwidth}
		\vspace{-4\baselineskip}
		\includegraphics[width=0.9\textwidth]{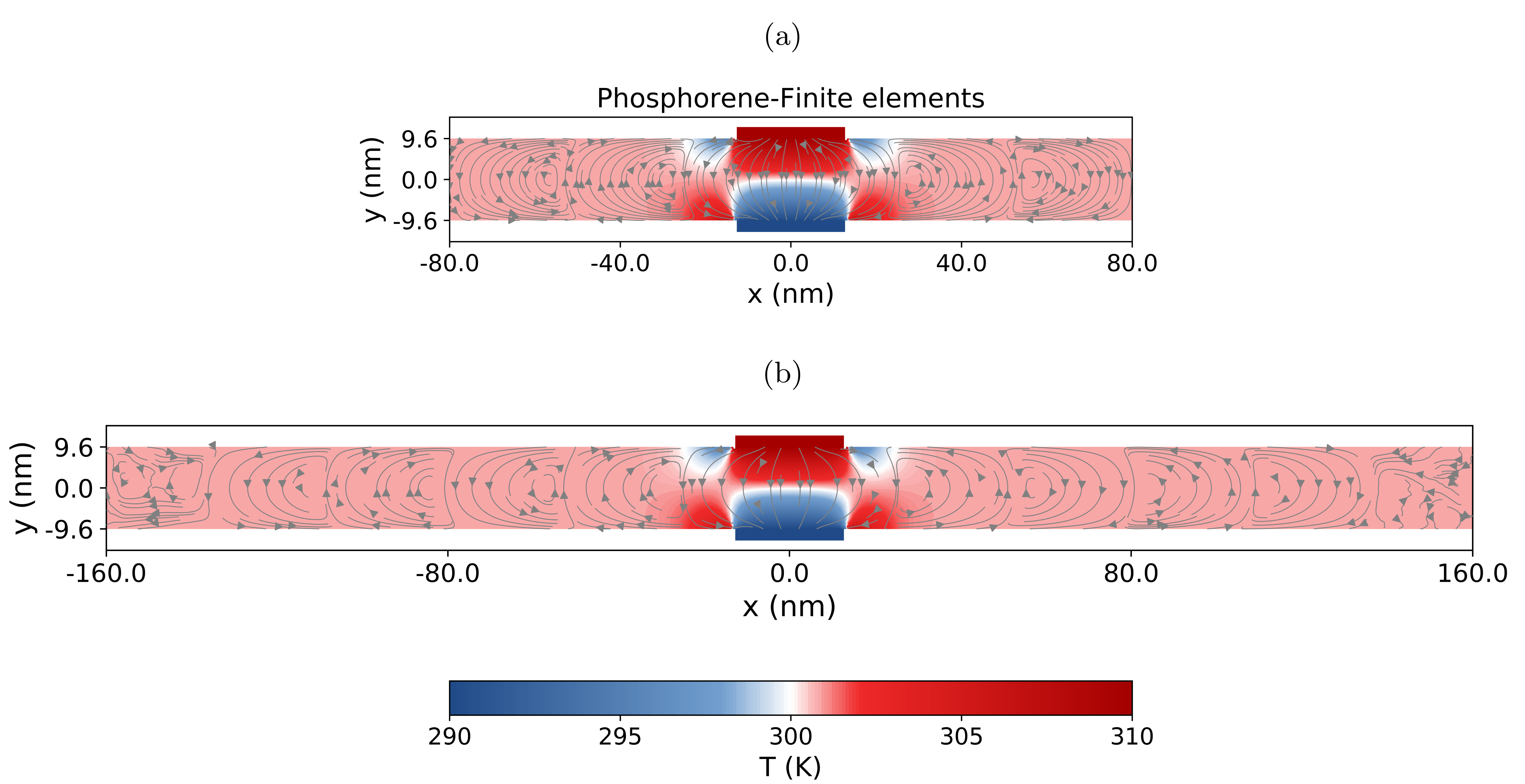}
	\end{subfigure}
	\caption{The finite elements solution of the hydrodynamic equation for two phosphorene-based Levitov configurations with $W_{\mathrm{reservoir}}=$ \SI{25}{\nano\meter}, $W_{\mathrm{device}}=$ \SI{160}{\nano\meter}\,(a)\,/\,\SI{320}{\nano\meter}\,(b) and $H=$ \SI{19.2}{\nano\meter}, $T_{\mathrm{HOT}}=$ \SI{310}{\kelvin} , $T_{\mathrm{COLD}}=$ \SI{290}{\kelvin},		 $\kappa_{\mathrm{AC}}=$\SI{10.47}{\watt\per\meter\per\kelvin}\,(a)\,/\,\SI{11.72}{\watt\per\meter\per\kelvin}\,(b), $\kappa_{\mathrm{ZZ}}=$\SI{17.15}{\watt\per\meter\per\kelvin}\,(a)\,/\,\SI{21.58}{\watt\per\meter\per\kelvin}\,(b),
		$\ell_{\mathrm{AC}}=$\SI{10.81}{\nm}\,(a)\,/\,\SI{13.30}{\nm}\,(b), and $\ell_{\mathrm{ZZ}}=$\SI{9.04}{\nm}\,(a)\,/\,\SI{11.60}{\nm}\,(b).
		The temperature profile is given in a blue-red colormap together with the heat flux isolines (gray). The dark-scarlet red and dark-sky blue boxes at the top and bottom are the hot and cold isothermal reservoirs, respectively.}
	\label{Fig:VORTEXPHOSPHORENE}
\end{figure*}

On top of that, and taking into account the role of boundary scattering on vortex formation pointed out by Shang~\textit{et al.}~\cite{Shang2020SR}, we also provide the PBTE solution without intrinsic scattering (i.e. three-phonon and isotopic scattering), henceforth referred to as extrinsic (see the bottom panels of Figs.~\ref{Fig:PbalDev} and~\ref{Fig:GPHbalDev}). From those results, it is clear that the boundary scattering is the only mechanism leading to vortex formation, as the other proposed contributing factor, the phase of phonons, is disregarded by the PBTE~\cite{ZhangAPLMat2021}. This last statement agrees with Zhang \textit{et al.}'s observations~\cite{ZhangIJHMT2021}: if allowed by boundary scattering, i.e. system geometry, vortices will form, being protected from destruction by the absence or scarcity of intrinsic resistive scattering processes. Furthermore, phosphorene extrinsic results show a curious feature, two secondary vortices, which are found neither in the RTA nor in bRTA results. We need to stress that the flux magnitude of such secondary vortices is well below---$\sim 10^4$ times smaller than the maximum, see Fig.~\ref{Fig:DoubleVortexbP}---the error introduced by the scattering algorithms~\cite{PeraudARHT2014}, so the observed flux in that area for the RTA and bRTA is nothing but noise-masked flux. To check for the existence of those secondary vortices, we resorted to the hydrodynamic equation using the cumulative quantities for the given device dimension as equation parameters; the result is given in Fig.~\ref{Fig:VORTEXPHOSPHORENE:small}.

Although there exist differences between this solution of the hydrodynamic equation and the MC simulations, those can be attributed to the facts that cumulative variables only capture the qualitative effects of size and that Eq.~\ref{Eq:HYDRO} does not capture ballistic effects present in the device. Moreover, we recall that the calculation of the $\ell$ values used here was conducted under the assumption of isotropy. Notwithstanding all these differences, this result together with extrinsic simulation strongly suggests the presence of additional vortices hidden by noise in Fig.~\ref{Fig:PbalDev}.


\begin{figure}[ht!]
	\centering
	\includegraphics[width=0.45\textwidth]{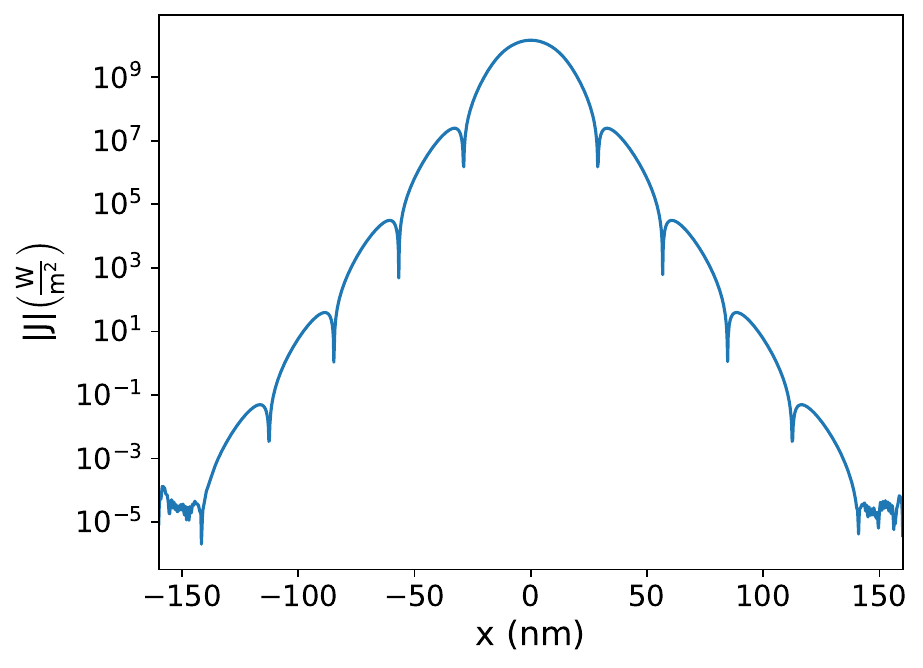}
	\caption{Cut of heat flux magnitude at Fig.~\ref{Fig:VORTEXPHOSPHORENE:large} for $y=$\SI{0}{\nano\meter} as function of $x$.}
	\label{Fig:VORTEXPHOSPHORENE:large-cut}
\end{figure}

\begin{figure}[ht!]
	\centering
	\includegraphics[width=0.45\textwidth]{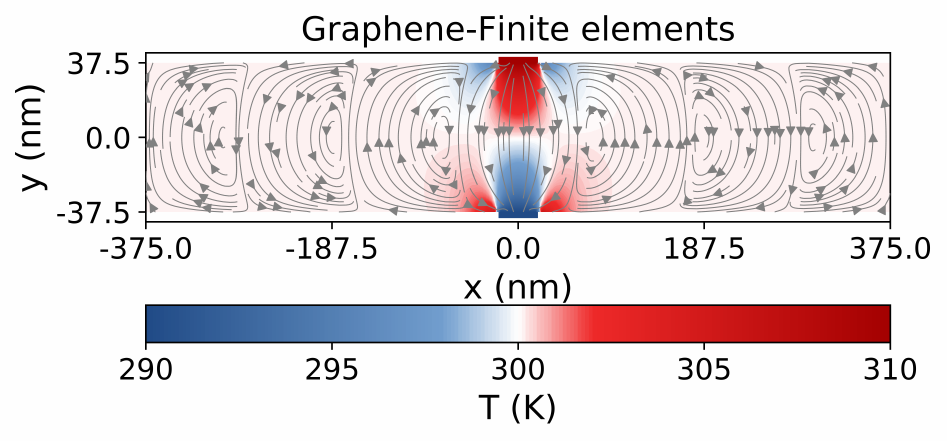}
	\caption{The finite elements solution of the hydrodynamic equation for a graphene-based Levitov configuration with $W_{\mathrm{reservoir}}=$ \SI{37.5}{\nano\meter}, $W_{\mathrm{device}}=$ \SI{750}{\nano\meter} and $H=$ \SI{75}{\nano\meter}, $T_{\mathrm{HOT}}=$ \SI{310}{\kelvin} , $T_{\mathrm{COLD}}=$ \SI{290}{\kelvin}, $\kappa=$\SI{374.08}{\watt\per\meter\per\kelvin}, and $\ell=$\SI{34.73}{\nm}. The temperature profile is given in a blue-red colormap together with the heat flux isolines (gray). The dark-scarlet red and dark-sky blue boxes at the top and bottom are the hot and cold isothermal reservoirs, respectively.}
	\label{Fig:DoubleVortexFD3}
\end{figure}

We also provide evidence of the existence of ternary and higher order vortices by solving the hydrodynamic equation for a wider phosphorene device (see Fig.~\ref{Fig:VORTEXPHOSPHORENE:large}). Interestingly, we observe from those results that each vortex is approximately 100 times smaller in magnitude than the previous one (see Fig.~\ref{Fig:VORTEXPHOSPHORENE:large-cut}) in line with MC simulation (Fig.~\ref{Fig:DoubleVortexbP}). We note that the noise at both lateral edges of the larger phosphorene device is due to the finite numerical precision in the simulations. Additionally, we complement those results with the hydrodynamic equation solutions for a larger graphene device (see Fig.~\ref{Fig:DoubleVortexFD3}), showing that the train of vortices is a general solution, and not a particular solution of phosphorene due to its properties. Indeed, the observation of a vortex train in wider devices ($W_{\mathrm{device}} \gg \lambda, W_{\mathrm{reservoir}}$) provides further information about the mechanism behind vorticity, suggesting that boundary scattering over lateral walls plays a secondary role in vortex generation at most.

\begin{figure*}[ht!]
	\centering
	\includegraphics[width=0.9\textwidth]{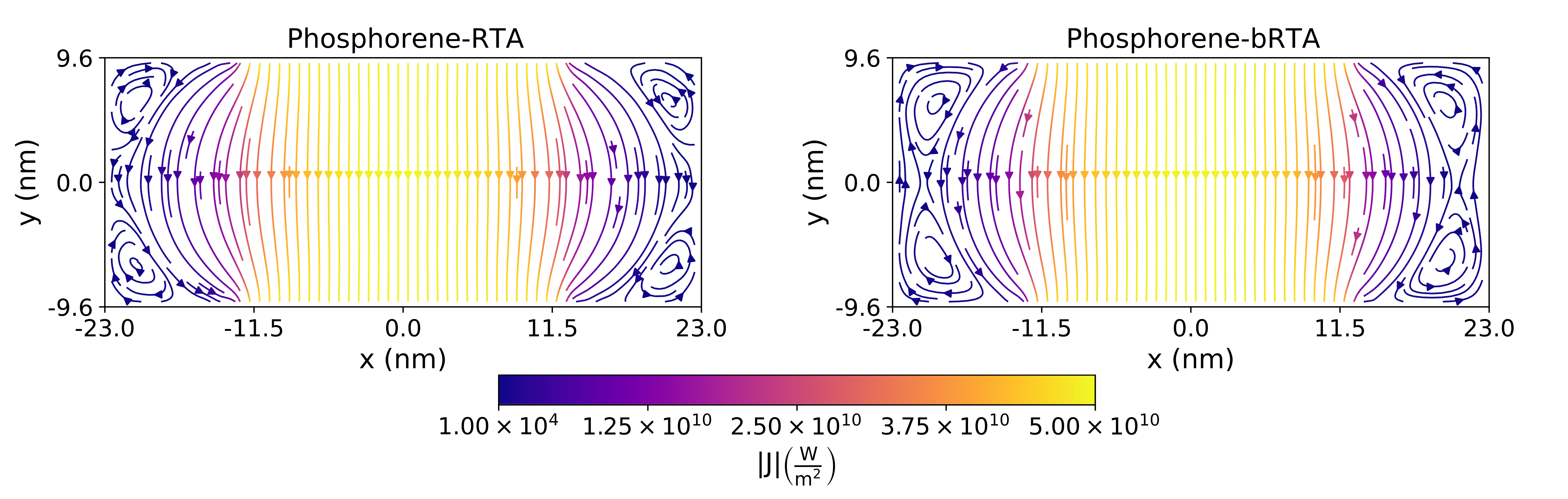}
	\caption{The RTA (left) and bRTA (right) steady-state heat fluxes for a limiting case phosphorene-based Levitov configuration with $W_{\mathrm{reservoir}}=$ \SI{25}{\nano\meter}, $W_{\mathrm{device}}=$ \SI{46}{\nano\meter} and $H=$ \SI{19.2}{\nano\meter}, $T_{\mathrm{HOT}}=$ \SI{310}{\kelvin} and $T_{\mathrm{COLD}}=$ \SI{290}{\kelvin}.}
	\label{Fig:limit}
\end{figure*}

In addition to providing information about the mechanism behind the vortex formation, the similarity between the RTA and bRTA fluxes suggests that because of boundary scattering dominance in the (quasi)ballistic regime, a proper description of the intrinsic scattering operator might be unnecessary to obtain a correct qualitative behavior of the flux. This contrasts with the findings in Ref.~\cite{myself}, in which different values of non-local length, the defining parameter of vorticity, for the RTA and bRTA were found in nanoribbons of sizes similar to those found in our devices. In view of such discrepancies, we investigated the existence of limiting cases, namely those cases in which a small difference in $\ell$ might lead to different vorticity. In Fig.~\ref{Fig:limit} we can see an example of such a limiting case for phosphorene, confirming the importance of a proper treatment of scattering even in the quasiballistic regime for a proper description of vortices. There, it can be observed that while in the RTA there are four different vortices, in the bRTA case there are still two different vortices. Moreover, those results also show vortex damping due to the lateral wall, demonstrating that a minimum distance between lateral walls and the reservoir ends is necessary for vortices to form~\cite{ZhangIJHMT2021}. This is not surprising, as in the limit of thermal reservoirs occupying all top and bottom edges one must recover a Poiseuille flow.

\subsubsection{\label{sec:Results:Ballistic:VortexFormation}Vortex formation}

\begin{figure*}[ht!]
	\centering
	\includegraphics[width=0.9\textwidth]{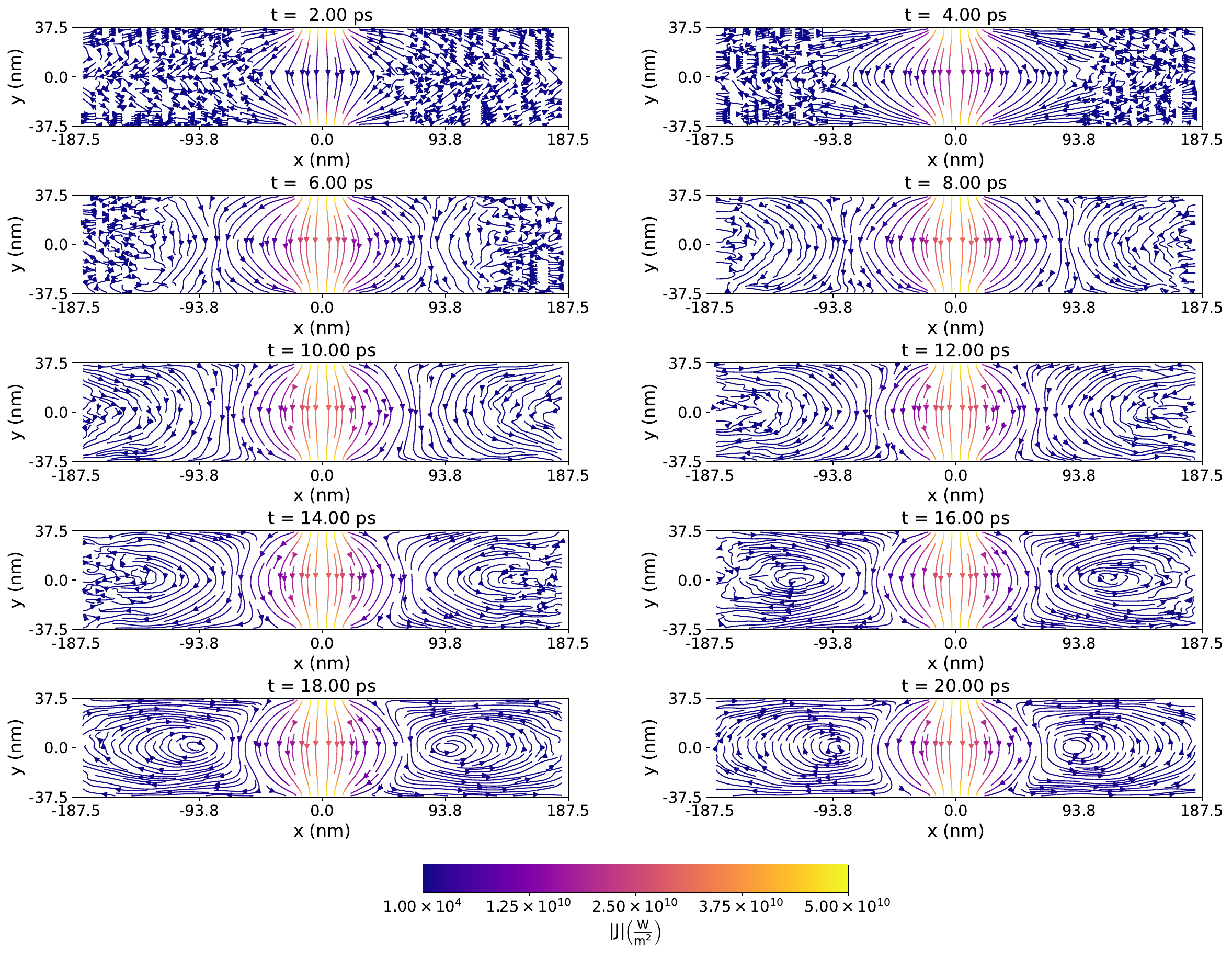}
	\caption{The bRTA heat flux at different times for a graphene-based Levitov configuration with $W_{\mathrm{reservoir}}=$ \SI{37.5}{\nano\meter}, $W_{\mathrm{device}}=$ \SI{375}{\nano\meter} and $H=$ \SI{75}{\nano\meter}, $T_{\mathrm{HOT}}=$ \SI{310}{\kelvin} and $T_{\mathrm{COLD}}=$ \SI{290}{\kelvin}, initially at \SI{300}{\kelvin}.}
	\label{Fig:VortexFormation}
\end{figure*}

To obtain a better understanding about how boundary scattering creates vortices, we look into how they are formed. For that purpose, we provide temporal snapshots showing the vortex formation within the bRTA framework, starting from a graphene-based Levitov configuration with $W_{\mathrm{reservoir}}=$ \SI{37.5}{\nano\meter}, $W_{\mathrm{device}}=$ \SI{375}{\nano\meter} and $H=$ \SI{75}{\nano\meter}, $T_{\mathrm{HOT}}=$ \SI{310}{\kelvin} and $T_{\mathrm{COLD}}=$ \SI{290}{\kelvin}, initially at \SI{300}{\kelvin} in Fig.~\ref{Fig:VortexFormation}. For a video of the time evolution of the heat flux and temperature see the \texttt{flux\_short.mp4} and \texttt{temperature\_short.mp4} videos in the Supplemental Material~\cite{SM}, respectively.

These results show that the formation of vortices begins at the sources, and how they advance with time towards the lateral edges until their closure. In further detail, the phonons arrive at the opposite vertical wall and slide laterally along it until they are stopped by boundary scattering, or collide with the lateral walls, closing the vortex in both cases. In the former, we recall that walls act as phonon sources due to diffusive scattering, and thus they reemit a small part of those incident phonons, which again slide laterally along the edges until they are stopped by edge friction or the collision with the lateral wall forming in that way additional vortices.

\begin{figure*}[ht!]
	\centering
	\includegraphics[width=0.9\textwidth]{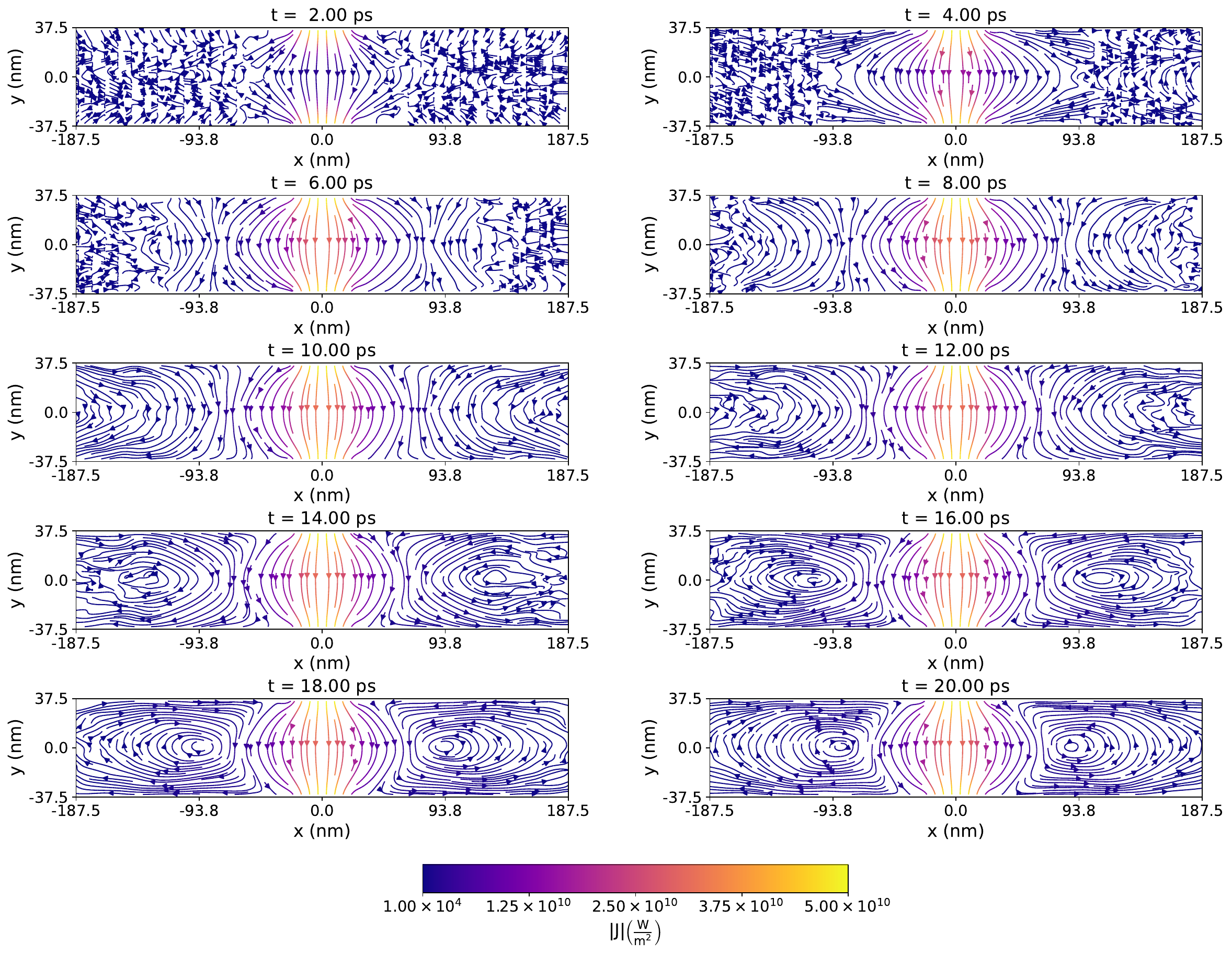}
	\caption{The bRTA heat flux at different times for a large graphene-based Levitov configuration with $W_{\mathrm{reservoir}}=$ \SI{37.5}{\nano\meter}, $W_{\mathrm{device}}=$ \SI{3.75}{\micro\meter} and $H=$ \SI{75}{\nano\meter}, $T_{\mathrm{HOT}}=$ \SI{310}{\kelvin} and $T_{\mathrm{COLD}}=$ \SI{290}{\kelvin}, initially at \SI{300}{\kelvin}. Only the central part of the device is depicted as most of the device is empty of phonons.}
	\label{Fig:VortexFormationLARGE}
\end{figure*}

Thus, such behavior supports the importance of top and bottom boundary scattering, as for larger devices vortices can start forming before a significant number of phonons can scatter at lateral borders. This latter statement is further confirmed by larger device simulations in which no phonon arrived at a lateral wall but the vortices are formed (see Fig.~\ref{Fig:VortexFormationLARGE}, for an animated version of the time evolution of the heat flux and temperature, see the \texttt{flux\_large.mp4} and \texttt{temperature\_larger.mp4} videos in the Supplemental Material~\cite{SM}, respectively).

\subsection{\label{sec:Results:NonBallistic}Non-ballistic devices}

Here we provide results for less ballistic devices than
those of the previous section, i.e.\ $H\gtrsim\lambda$. As aforementioned, owing to the differences in the mean free path and non-local length between graphene and phosphorene, we expect both to show rather distinct vorticity.

\subsubsection{Phosphorene}

Devices out of the ballistic regime are in principle much more interesting in the case of phosphorene, as the non-local length is slightly larger than the mean free path, therefore allowing for hydrodynamic features outside the ballistic regime. Moreover, based on the fit of $\ell$ to Monte Carlo results found in Ref.~\cite{myself}, we expect its value to be larger than the calculated one, thus allowing for vortices and negative thermal resistances in devices with $H$ larger than what would be expected from the results in Table~\ref{Tab:BULK}.

\begin{figure}[ht!]
	\centering
	\includegraphics[width=0.45\textwidth]{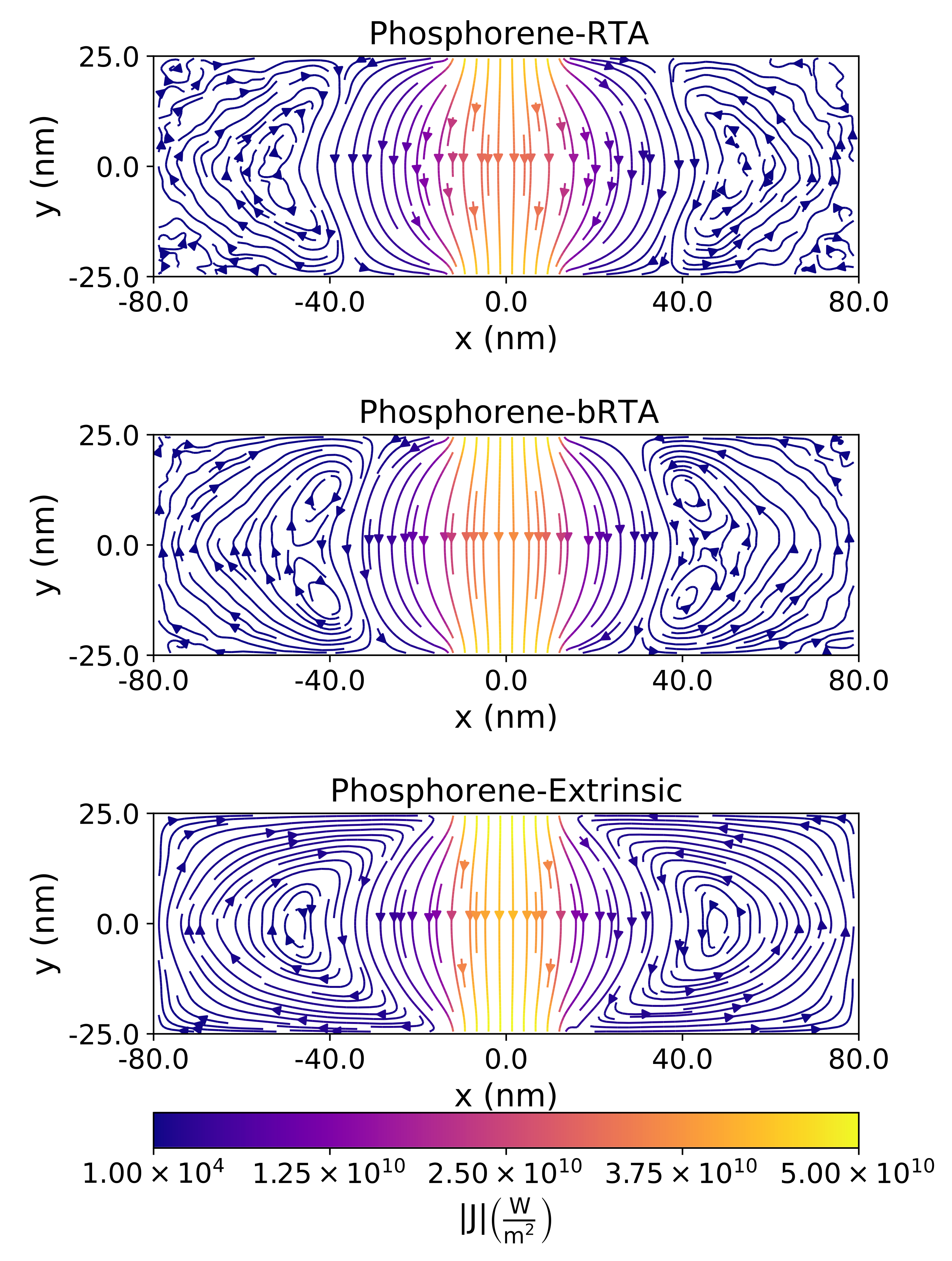}
	\caption{The RTA (top), bRTA (middle), and extrinsic (bottom) steady-state heat fluxes for phosphorene-based Levitov configuration with $W_{\mathrm{reservoir}}=$ \SI{25}{\nano\meter}, $W_{\mathrm{device}}=$ \SI{160}{\nano\meter} and $H=$ \SI{50}{\nano\meter}, $T_{\mathrm{HOT}}=$ \SI{310}{\kelvin} and $T_{\mathrm{COLD}}=$ \SI{290}{\kelvin}.}
	\label{Fig:P50}
\end{figure}

\begin{figure}[ht!]
	\centering
	\includegraphics[width=0.45\textwidth]{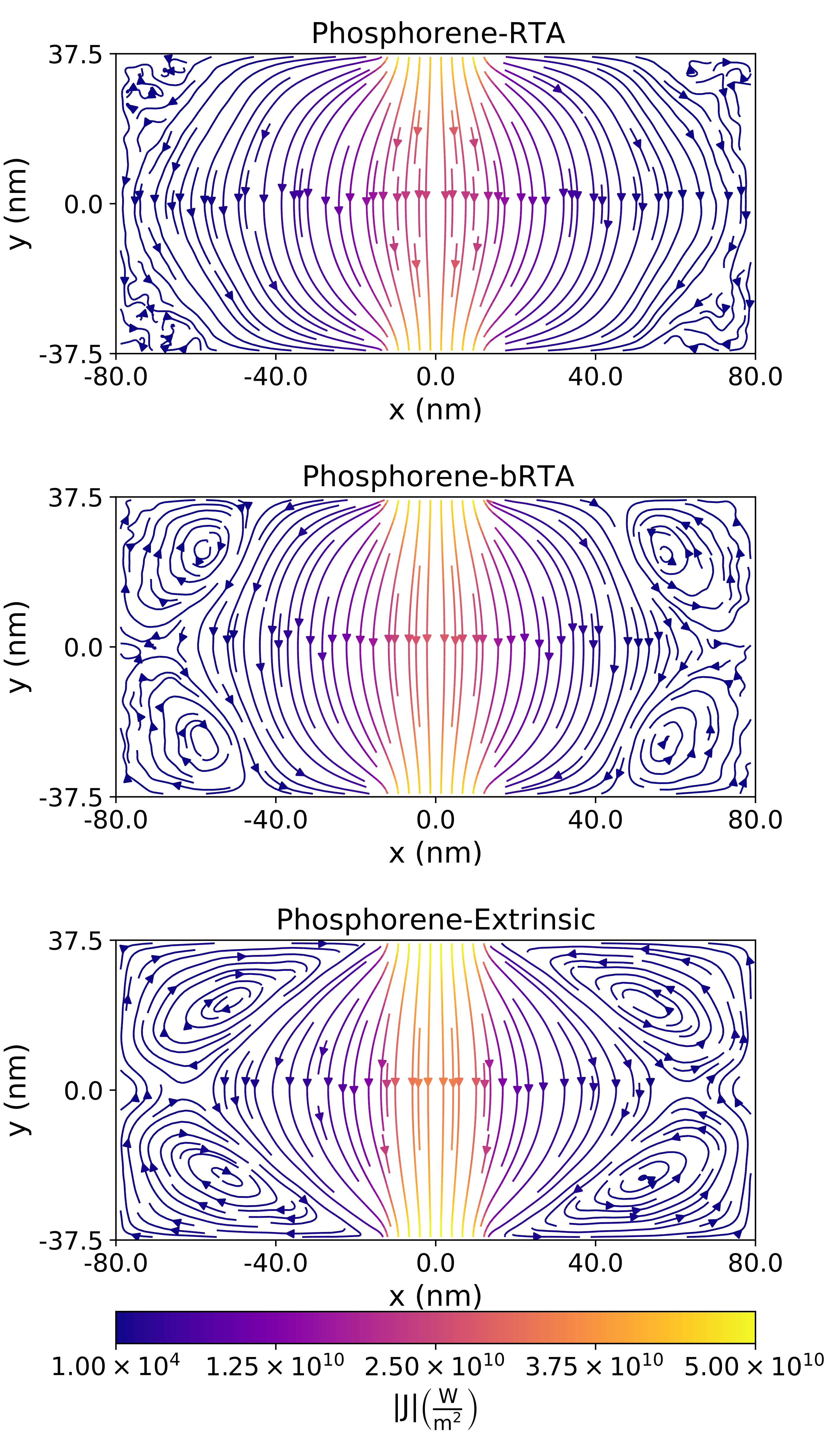}
	\caption{The RTA (top), bRTA (middle), and extrinsic (bottom) steady-state heat fluxes for phosphorene-based Levitov configuration with $W_{\mathrm{reservoir}}=$ \SI{25}{\nano\meter}, $W_{\mathrm{device}}=$ \SI{160}{\nano\meter} and $H=$ \SI{75}{\nano\meter}, $T_{\mathrm{HOT}}=$ \SI{310}{\kelvin} and $T_{\mathrm{COLD}}=$ \SI{290}{\kelvin}.}
	\label{Fig:P75}
\end{figure}

\begin{figure}[ht!]
	\centering
	\includegraphics[width=0.45\textwidth]{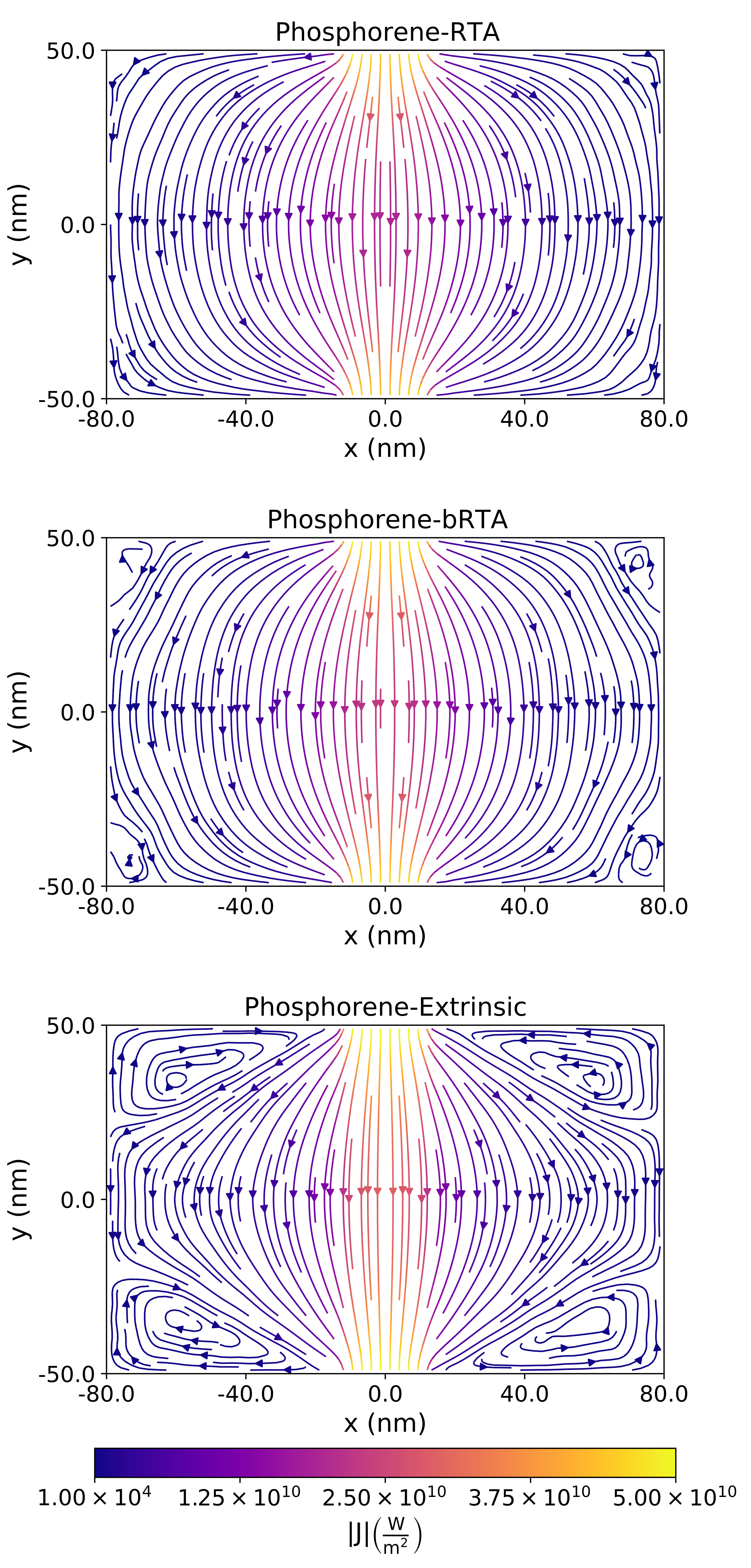}
	\caption{The RTA (top), bRTA (middle), and extrinsic (bottom) steady-state heat fluxes for phosphorene-based Levitov configuration with $W_{\mathrm{reservoir}}=$ \SI{25}{\nano\meter}, $W_{\mathrm{device}}=$ \SI{160}{\nano\meter} and $H=$ \SI{100}{\nano\meter}, $T_{\mathrm{HOT}}=$ \SI{310}{\kelvin} and $T_{\mathrm{COLD}}=$ \SI{290}{\kelvin}.}
	\label{Fig:P100}
\end{figure}

\begin{figure}[ht!]
	\centering
	\begin{subfigure}[b]{0.45\textwidth}
		\captionsetup{labelformat=empty}
		\caption{}
		\label{Fig:T50}
	\end{subfigure}
	\begin{subfigure}[b]{0.45\textwidth}
		\captionsetup{labelformat=empty}
		\caption{}
		\label{Fig:T75}
	\end{subfigure}
	\begin{subfigure}[b]{0.45\textwidth}
		\captionsetup{labelformat=empty}
		\caption{}
		\label{Fig:T100}
	\end{subfigure}
    \begin{subfigure}[b]{0.45\textwidth}
    	\vspace{-6\baselineskip}
    	\includegraphics[width=\textwidth]{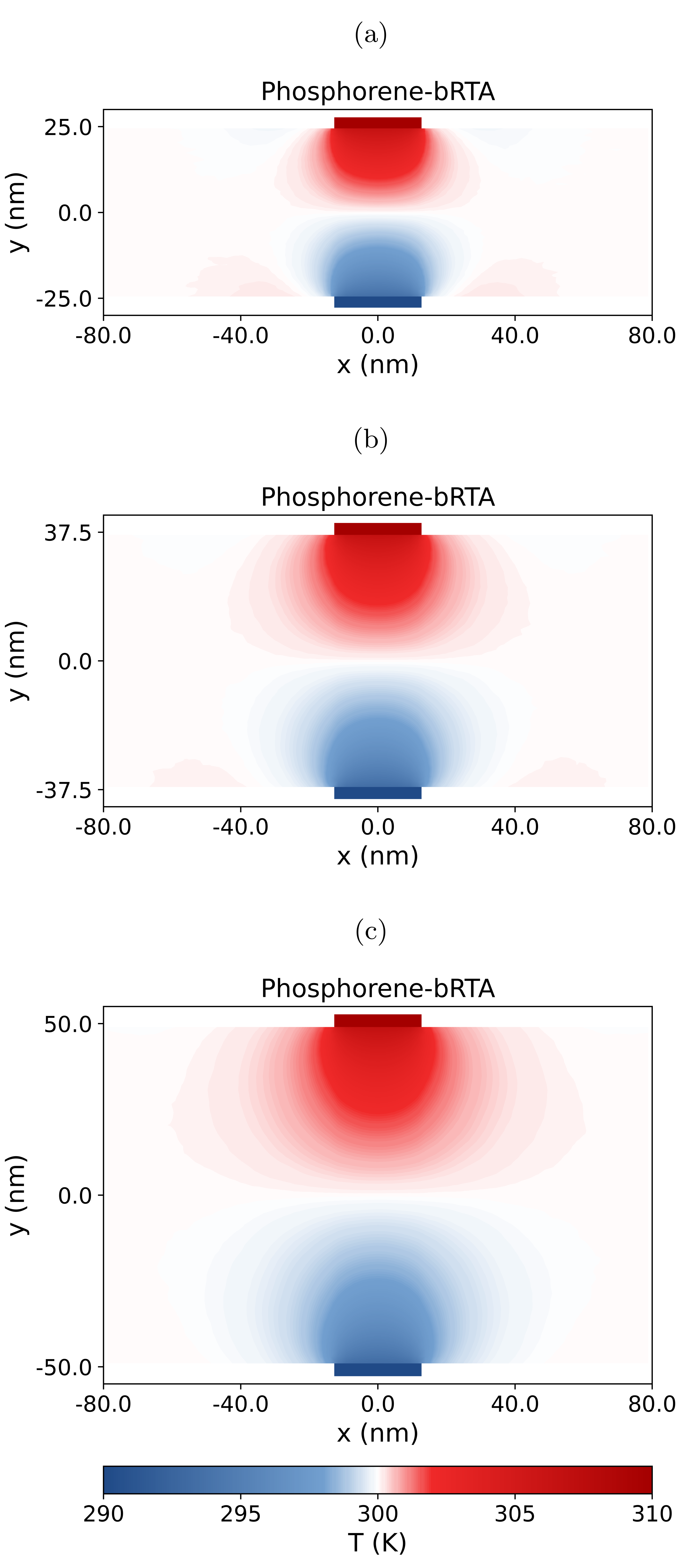}
    \end{subfigure}
	\caption{The bRTA steady-state temperature profile for several phosphorene-based Levitov configurations with $W_{\mathrm{reservoir}}=$ \SI{25}{\nano\meter}, $W_{\mathrm{device}}=$ \SI{160}{\nano\meter} and $H=$ \SI{50}{\nano\meter}\,(a)\,/\,\SI{75}{\nano\meter}\,(b)\,/\,\SI{100}{\nano\meter}\,(c), $T_{\mathrm{HOT}}=$ \SI{310}{\kelvin} and $T_{\mathrm{COLD}}=$ \SI{290}{\kelvin}. The dark-scarlet red and dark-sky blue boxes at the top and bottom are the hot and cold isothermal reservoirs, respectively.}
	\label{Fig:Tall}
\end{figure}

Because of this, we have simulated three different heights: 50, 75, and \SI{100}{\nano\meter}, the RTA and bRTA steady-state heat fluxes of which are shown in Figs.~\ref{Fig:P50}-\ref{Fig:P100}. These results are to be compared to the extrinsic-only solution for the same set of heights (see the extrinsic cases at the bottom of Figs~\ref{Fig:P50}-\ref{Fig:P100}) to highlight the effect of intrinsic scattering on the steady-state properties. Additionally, we provide the bRTA thermal profiles for all heights in Figs.~\ref{Fig:T50}-\ref{Fig:T100}.
In smaller devices, we observe that device geometry allows for the formation of two vortices (see the extrinsic case at the bottom of Fig.~\ref{Fig:P50}), which are present even when intrinsic scattering is added. Interestingly, if we set aside the expected magnitude differences between all the treatments of the scattering operator, we can observe (Fig.~\ref{Fig:P50}) subtle differences in the vortex profiles. While within the extrinsic treatment we can see two clean vortices, the addition of intrinsic scattering through the RTA causes those vortices to start dissipating at the device's corners. However, the proper treatment of scattering leads to slightly different alterations, as such dissipation of vorticity is less pronounced, while the RTA incipient fracture of principal vortices into two becomes clearer in the bRTA case. Therefore, the full linearized scattering operator is necessary to obtain an appropriate description of vorticity.

The 75 and \SI{100}{\nano\meter} results (see Figs.~\ref{Fig:P75} and~\ref{Fig:P100}) further support this latter statement. In both cases vorticity is geometrically permitted (see the extrinsic cases at the bottom of Figs~\ref{Fig:P75} and \ref{Fig:P100}) but suppressed in the RTA case, which shows an almost perfect diffusive profile, especially for the \SI{100}{\nano\metre} device. On the other hand, the correct description of the scattering operator allows the vortices to survive in both cases. This fact is nothing but a consequence of the RTA deeming all three-phonon processes as resistive, which translates into lower $\ell$ values. Indeed, we expect the flux direction to be qualitatively similar for devices much larger than $\ell$, as they tend to have diffusive profiles---i.e. profiles compatible with Fourier's law---in which the only difference between the RTA and bRTA is the flux magnitude due to the thermal conductivity differences. 

Regarding thermal resistance, we can observe a low-magnitude negative resistance, when compared to ballistic devices, for the 50 device (Figs.~\ref{Fig:P50} and \ref{Fig:T50}), and even a lower one in the \SI{75}{\nano\meter} one (Figs.~\ref{Fig:P75} and \ref{Fig:T75}). For the larger device (i.e. \SI{100}{\nano\meter}, Figs.~\ref{Fig:P100} and \ref{Fig:T100}) no significant negative thermal resistance is observed, indicating (in line with the flux results) a weakening of hydrodynamic regime with increasing $H$. This is a consequence of the phonon distribution losing its inertia---i.e. preventing the effect of boundary scattering---while traveling due to intrinsic resistive scattering, up until the point that for larger distances ($H$) it is completely lost before arriving to the opposite side, and thus one arrives at the Fourier's diffusive regime.

\subsubsection{Graphene}

\begin{figure}[ht!]
	\centering
	\includegraphics[width=0.45\textwidth]{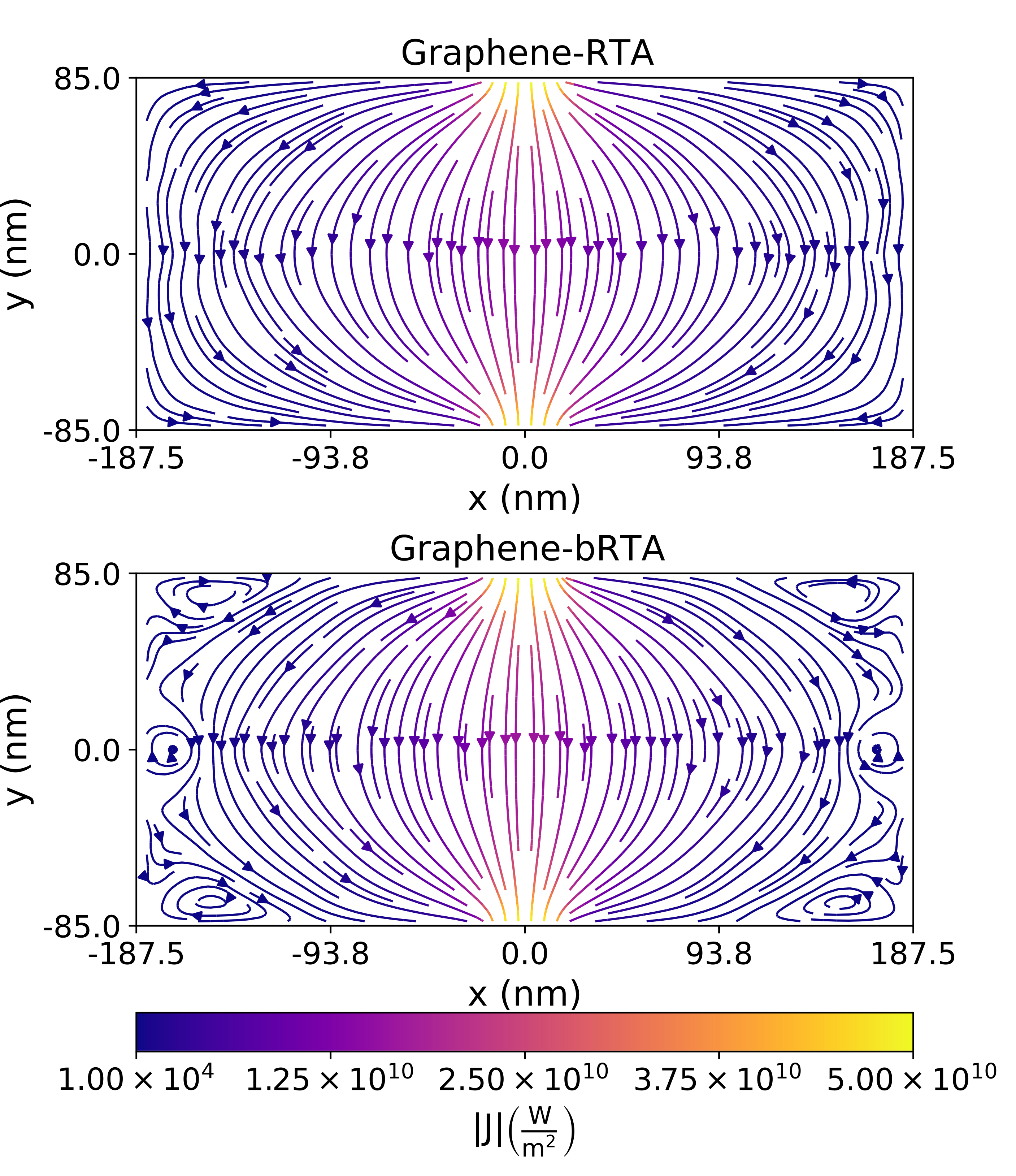}
	\caption{The RTA (top) and bRTA (bottom) steady-state heat fluxes for graphene-based Levitov configuration with $W_{\mathrm{reservoir}}=$ \SI{37.5}{\nano\meter}, $W_{\mathrm{device}}=$ \SI{375}{\nano\meter} and $H=$ \SI{170}{\nano\meter}, $T_{\mathrm{HOT}}=$ \SI{310}{\kelvin} and $T_{\mathrm{COLD}}=$ \SI{290}{\kelvin}.}
	\label{Fig:GPHnoBAL}
\end{figure}

\begin{figure}[ht!]
	\centering
	\includegraphics[width=0.45\textwidth]{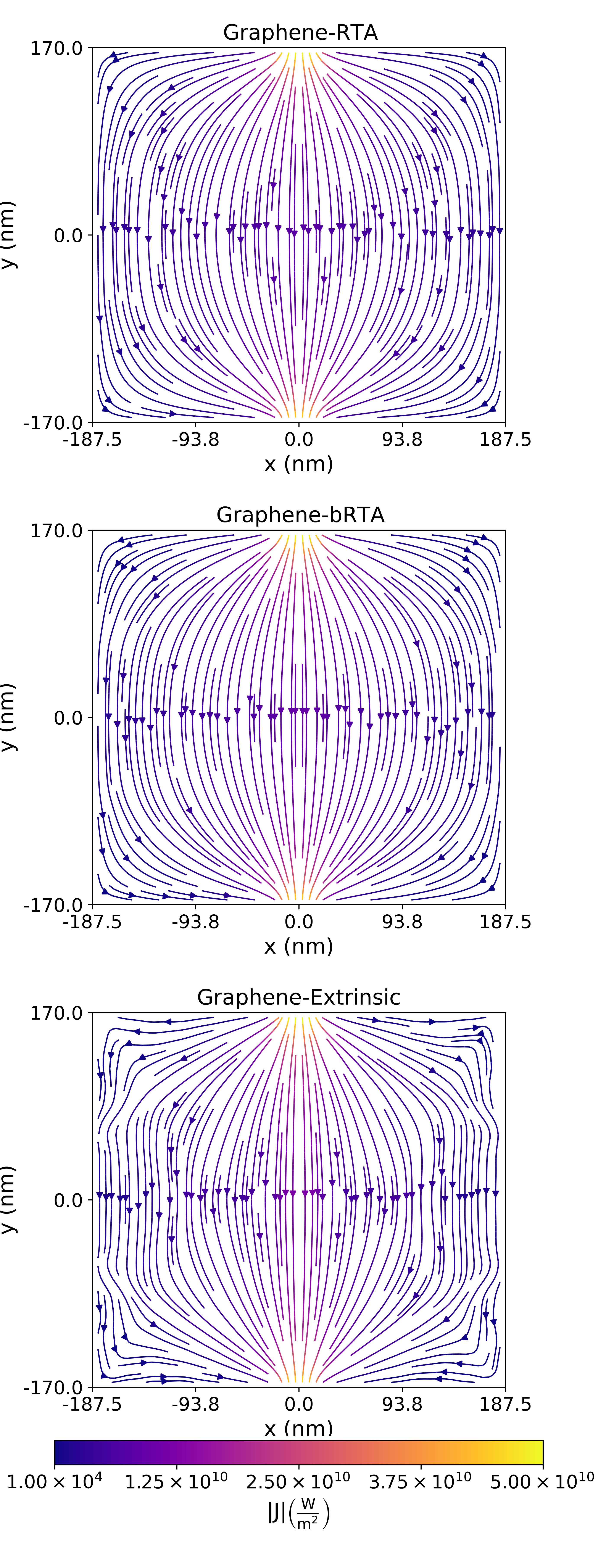}
	\caption{The RTA (top), bRTA (middle) and extrinsic (bottom) steady-state heat fluxes for graphene-based Levitov configuration with $W_{\mathrm{reservoir}}=$ \SI{37.5}{\nano\meter}, $W_{\mathrm{device}}=$ \SI{375}{\nano\meter} and $H=$ \SI{340}{\nano\meter}, $T_{\mathrm{HOT}}=$ \SI{310}{\kelvin} and $T_{\mathrm{COLD}}=$ \SI{290}{\kelvin}.}
	\label{Fig:GPHnoBALLARGE}
\end{figure}

The graphene RTA and bRTA steady-state heat fluxes are depicted in Fig.~\ref{Fig:GPHnoBAL}, showing a clear diffusive (Fourier-like) profile in the RTA case. In contrast, some vorticity can be observed in the bRTA results, indicating a clear underestimation of non-local length---i.e. the ability of the distribution to keep its inertia---because of the RTA limitations. Therefore, it is indeed possible to have some small vorticity out of the ballistic regime. 
When incrementing the height, we observe that vortices cannot form even without intrinsic scattering (see the extrinsic case at the bottom of Fig.~\ref{Fig:GPHnoBALLARGE}) although some non-local effects appear in the flux, namely the curvature of the isolines near the corners of the extrinsic case in Fig.~\ref{Fig:GPHnoBALLARGE}. However, such hydrodynamic effects are not present neither in the RTA nor bRTA (see the top and middle panels of Fig.~\ref{Fig:GPHnoBALLARGE}) due to the intrinsic scattering effect (i.e. $\ell\ll H$), showing both a Fourier-like profile. 

\section{\label{sec:Conclusions} Summary and conclusions}

In this work, we have presented Monte Carlo simulations of phonon transport with a full linearized \textit{ab initio} scattering operator, leading to the manifestation of hydrodynamic signatures such as vorticity and non-monotonic temperature profiles in planar graphene and phosphorene device geometries. We find such signatures at room temperature in quasiballistic and non-ballistic regimes, in contradiction with the predictions of classical hydrodynamic models, which required a predominance of normal scattering, but in agreement with more recent and generalized models which find the signatures even when all scattering is resistive. In contrast to previous observations, we demonstrate that there is a shared mechanism among all transport regimes originating the hydrodynamic signatures, namely boundary scattering. Indeed, we show that in our devices, undergoing vertical transport, boundary scattering at horizontal edges is the leading mechanism behind vortex formation, proving through the solution of the hydrodynamic equation that an increment in the horizontal dimension only leads to the formation of more convection cells instead of vortex (hydrodynamics) destruction. We have illustrated the critical importance of the device dimensions, especially the distance between sources and the opposite horizontal walls (i.e. the ones causing the vortices), in determining the presence or absence of hydrodynamic effects, opposite to previous considerations based only on the relative importance of normal vs.\ resistive scattering. Moreover, we have demonstrated that having a proper description of the intrinsic scattering using the full linearized scattering operator will, through more accurate values of $\ell$, result in qualitative consequences, showing that the RTA can lead to an incorrect number of vortices or no vorticity at all in cases where it should be present.

Finally, based on our calculations, we point out that experimental observation of hydrodynamic signatures at room temperature would be favored by choosing a small graphene sample under \SI{100}{\nano\meter} in height, a finite source and sink as narrow as possible and a device width at least three times larger than the source/sink widths.

\section*{Acknowledgements}
X.\,C. acknowledges discussions with Prof.\ X.~Alvarez. M.\,R.-M.\ and X.\,C. acknowledge financial support by Spain's Ministerio de Ciencia, Innovaci\'on y Universidades under Grant No.\,RTI2018-097876-B-C21 (MCIU/AEI/FEDER, UE), and the EU Horizon2020 research and innovation program under grant GrapheneCore3 881603.
M.\,R.-M.\ acknowledges financial support by the Ministerio de Educaci\'on, Cultura y Deporte programme of Formaci\'on de Profesorado Universitario under Grant No. FPU2016/02565 and Ministerio de Universidades under grant EST19/00655, as well as the kind hospitality of the Institute of Materials Chemistry, TU Wien.

\bibliography{ref}

\begin{thebibliography}{62}%
\makeatletter
\providecommand \@ifxundefined [1]{%
 \@ifx{#1\undefined}
}%
\providecommand \@ifnum [1]{%
 \ifnum #1\expandafter \@firstoftwo
 \else \expandafter \@secondoftwo
 \fi
}%
\providecommand \@ifx [1]{%
 \ifx #1\expandafter \@firstoftwo
 \else \expandafter \@secondoftwo
 \fi
}%
\providecommand \natexlab [1]{#1}%
\providecommand \enquote  [1]{``#1''}%
\providecommand \bibnamefont  [1]{#1}%
\providecommand \bibfnamefont [1]{#1}%
\providecommand \citenamefont [1]{#1}%
\providecommand \href@noop [0]{\@secondoftwo}%
\providecommand \href [0]{\begingroup \@sanitize@url \@href}%
\providecommand \@href[1]{\@@startlink{#1}\@@href}%
\providecommand \@@href[1]{\endgroup#1\@@endlink}%
\providecommand \@sanitize@url [0]{\catcode `\\12\catcode `\$12\catcode
  `\&12\catcode `\#12\catcode `\^12\catcode `\_12\catcode `\%12\relax}%
\providecommand \@@startlink[1]{}%
\providecommand \@@endlink[0]{}%
\providecommand \url  [0]{\begingroup\@sanitize@url \@url }%
\providecommand \@url [1]{\endgroup\@href {#1}{\urlprefix }}%
\providecommand \urlprefix  [0]{URL }%
\providecommand \Eprint [0]{\href }%
\providecommand \doibase [0]{http://dx.doi.org/}%
\providecommand \selectlanguage [0]{\@gobble}%
\providecommand \bibinfo  [0]{\@secondoftwo}%
\providecommand \bibfield  [0]{\@secondoftwo}%
\providecommand \translation [1]{[#1]}%
\providecommand \BibitemOpen [0]{}%
\providecommand \bibitemStop [0]{}%
\providecommand \bibitemNoStop [0]{.\EOS\space}%
\providecommand \EOS [0]{\spacefactor3000\relax}%
\providecommand \BibitemShut  [1]{\csname bibitem#1\endcsname}%
\let\auto@bib@innerbib\@empty
\bibitem [{\citenamefont {Moore}(1965)}]{Moores_law}%
  \BibitemOpen
  \bibfield  {author} {\bibinfo {author} {\bibfnamefont {G.~E.}\ \bibnamefont
  {Moore}},\ }\href@noop {} {\bibfield  {journal} {\bibinfo  {journal}
  {Electronics}\ }\textbf {\bibinfo {volume} {38}},\ \bibinfo {pages} {114}
  (\bibinfo {year} {1965})}\BibitemShut {NoStop}%
\bibitem [{\citenamefont {Moore}\ and\ \citenamefont
  {Shi}(2014)}]{MooreMatToday2014}%
  \BibitemOpen
  \bibfield  {author} {\bibinfo {author} {\bibfnamefont {A.~L.}\ \bibnamefont
  {Moore}}\ and\ \bibinfo {author} {\bibfnamefont {L.}~\bibnamefont {Shi}},\
  }\href {\doibase https://doi.org/10.1016/j.mattod.2014.04.003} {\bibfield
  {journal} {\bibinfo  {journal} {Mater. Today}\ }\textbf {\bibinfo {volume}
  {17}},\ \bibinfo {pages} {163} (\bibinfo {year} {2014})}\BibitemShut
  {NoStop}%
\bibitem [{\citenamefont {Minnich}\ \emph {et~al.}(2009)\citenamefont
  {Minnich}, \citenamefont {Dresselhaus}, \citenamefont {Ren},\ and\
  \citenamefont {Chen}}]{MinnichEES2009}%
  \BibitemOpen
  \bibfield  {author} {\bibinfo {author} {\bibfnamefont {A.~J.}\ \bibnamefont
  {Minnich}}, \bibinfo {author} {\bibfnamefont {M.~S.}\ \bibnamefont
  {Dresselhaus}}, \bibinfo {author} {\bibfnamefont {Z.~F.}\ \bibnamefont
  {Ren}}, \ and\ \bibinfo {author} {\bibfnamefont {G.}~\bibnamefont {Chen}},\
  }\href {\doibase 10.1039/B822664B} {\bibfield  {journal} {\bibinfo  {journal}
  {Energy Environ. Sci.}\ }\textbf {\bibinfo {volume} {2}},\ \bibinfo {pages}
  {466} (\bibinfo {year} {2009})}\BibitemShut {NoStop}%
\bibitem [{\citenamefont {Maldovan}(2013)}]{maldovanNature2013}%
  \BibitemOpen
  \bibfield  {author} {\bibinfo {author} {\bibfnamefont {M.}~\bibnamefont
  {Maldovan}},\ }\href@noop {} {\bibfield  {journal} {\bibinfo  {journal}
  {Nature}\ }\textbf {\bibinfo {volume} {503}},\ \bibinfo {pages} {209}
  (\bibinfo {year} {2013})}\BibitemShut {NoStop}%
\bibitem [{\citenamefont {Li}\ \emph {et~al.}(2018)\citenamefont {Li},
  \citenamefont {Song}, \citenamefont {Zhao}, \citenamefont {Yang},
  \citenamefont {Pastel}, \citenamefont {Xu}, \citenamefont {Jia},
  \citenamefont {Dai}, \citenamefont {Chen}, \citenamefont {Gong},
  \citenamefont {Jiang}, \citenamefont {Yao}, \citenamefont {Fan},
  \citenamefont {Yang}, \citenamefont {W\r{a}gberg}, \citenamefont {Yang},\
  and\ \citenamefont {Hu}}]{TianSciAdv2018}%
  \BibitemOpen
  \bibfield  {author} {\bibinfo {author} {\bibfnamefont {T.}~\bibnamefont
  {Li}}, \bibinfo {author} {\bibfnamefont {J.}~\bibnamefont {Song}}, \bibinfo
  {author} {\bibfnamefont {X.}~\bibnamefont {Zhao}}, \bibinfo {author}
  {\bibfnamefont {Z.}~\bibnamefont {Yang}}, \bibinfo {author} {\bibfnamefont
  {G.}~\bibnamefont {Pastel}}, \bibinfo {author} {\bibfnamefont
  {S.}~\bibnamefont {Xu}}, \bibinfo {author} {\bibfnamefont {C.}~\bibnamefont
  {Jia}}, \bibinfo {author} {\bibfnamefont {J.}~\bibnamefont {Dai}}, \bibinfo
  {author} {\bibfnamefont {C.}~\bibnamefont {Chen}}, \bibinfo {author}
  {\bibfnamefont {A.}~\bibnamefont {Gong}}, \bibinfo {author} {\bibfnamefont
  {F.}~\bibnamefont {Jiang}}, \bibinfo {author} {\bibfnamefont
  {Y.}~\bibnamefont {Yao}}, \bibinfo {author} {\bibfnamefont {T.}~\bibnamefont
  {Fan}}, \bibinfo {author} {\bibfnamefont {B.}~\bibnamefont {Yang}}, \bibinfo
  {author} {\bibfnamefont {L.}~\bibnamefont {W\r{a}gberg}}, \bibinfo {author}
  {\bibfnamefont {R.}~\bibnamefont {Yang}}, \ and\ \bibinfo {author}
  {\bibfnamefont {L.}~\bibnamefont {Hu}},\ }\href {\doibase
  10.1126/sciadv.aar3724} {\bibfield  {journal} {\bibinfo  {journal} {Sci.
  Adv.}\ }\textbf {\bibinfo {volume} {4}},\ \bibinfo {pages} {eaar3724}
  (\bibinfo {year} {2018})}\BibitemShut {NoStop}%
\bibitem [{\citenamefont {Pop}\ \emph {et~al.}(2006)\citenamefont {Pop},
  \citenamefont {Sinha},\ and\ \citenamefont {Goodson}}]{PopPIEEE2006}%
  \BibitemOpen
  \bibfield  {author} {\bibinfo {author} {\bibfnamefont {E.}~\bibnamefont
  {Pop}}, \bibinfo {author} {\bibfnamefont {S.}~\bibnamefont {Sinha}}, \ and\
  \bibinfo {author} {\bibfnamefont {K.}~\bibnamefont {Goodson}},\ }\href
  {\doibase 10.1109/JPROC.2006.879794} {\bibfield  {journal} {\bibinfo
  {journal} {Proc. IEEE}\ }\textbf {\bibinfo {volume} {94}},\ \bibinfo {pages}
  {1587} (\bibinfo {year} {2006})}\BibitemShut {NoStop}%
\bibitem [{\citenamefont {Macii}(2004)}]{macii_book}%
  \BibitemOpen
  \bibfield  {author} {\bibinfo {author} {\bibfnamefont {E.}~\bibnamefont
  {Macii}},\ }\href@noop {} {\emph {\bibinfo {title} {Ultra low-power
  electronics and design}}}\ (\bibinfo  {publisher} {Springer},\ \bibinfo
  {year} {2004})\BibitemShut {NoStop}%
\bibitem [{\citenamefont {Panda}\ \emph {et~al.}(2010)\citenamefont {Panda},
  \citenamefont {Silpa}, \citenamefont {Shrivastava},\ and\ \citenamefont
  {Gummidipudi}}]{panda_book}%
  \BibitemOpen
  \bibfield  {author} {\bibinfo {author} {\bibfnamefont {P.~R.}\ \bibnamefont
  {Panda}}, \bibinfo {author} {\bibfnamefont {B.}~\bibnamefont {Silpa}},
  \bibinfo {author} {\bibfnamefont {A.}~\bibnamefont {Shrivastava}}, \ and\
  \bibinfo {author} {\bibfnamefont {K.}~\bibnamefont {Gummidipudi}},\
  }\href@noop {} {\emph {\bibinfo {title} {Power-efficient system design}}}\
  (\bibinfo  {publisher} {Springer Science \& Business Media},\ \bibinfo {year}
  {2010})\BibitemShut {NoStop}%
\bibitem [{\citenamefont {Ross}(2008)}]{RossIEEESpec2008}%
  \BibitemOpen
  \bibfield  {author} {\bibinfo {author} {\bibfnamefont {P.~E.}\ \bibnamefont
  {Ross}},\ }\href@noop {} {\bibfield  {journal} {\bibinfo  {journal} {IEEE
  Spectr.}\ }\textbf {\bibinfo {volume} {45}},\ \bibinfo {pages} {72} (\bibinfo
  {year} {2008})}\BibitemShut {NoStop}%
\bibitem [{\citenamefont {Fourier}(1822)}]{Fourier1822}%
  \BibitemOpen
  \bibfield  {author} {\bibinfo {author} {\bibfnamefont {J.}~\bibnamefont
  {Fourier}},\ }\href {https://books.google.es/books?id=1TUVAAAAQAAJ} {\emph
  {\bibinfo {title} {Th{\'e}orie analytique de la chaleur}}}\ (\bibinfo
  {publisher} {Chez Firmin Didot, p{\`e}re et fils},\ \bibinfo {year}
  {1822})\BibitemShut {NoStop}%
\bibitem [{\citenamefont {Wilson}\ and\ \citenamefont
  {Cahill}(2014)}]{WilsonNC2014}%
  \BibitemOpen
  \bibfield  {author} {\bibinfo {author} {\bibfnamefont {R.~B.}\ \bibnamefont
  {Wilson}}\ and\ \bibinfo {author} {\bibfnamefont {D.~G.}\ \bibnamefont
  {Cahill}},\ }\href {\doibase 10.1038/ncomms6075} {\bibfield  {journal}
  {\bibinfo  {journal} {Nat. Commun.}\ }\textbf {\bibinfo {volume} {5}},\
  \bibinfo {pages} {5075} (\bibinfo {year} {2014})}\BibitemShut {NoStop}%
\bibitem [{\citenamefont {Ziabari}\ \emph {et~al.}(2018)\citenamefont
  {Ziabari}, \citenamefont {Torres}, \citenamefont {Vermeersch}, \citenamefont
  {Xuan}, \citenamefont {Cartoix{\`a}}, \citenamefont {Torell{\'o}},
  \citenamefont {Bahk}, \citenamefont {Koh}, \citenamefont {Parsa},
  \citenamefont {Ye}, \citenamefont {Alvarez},\ and\ \citenamefont
  {Shakouri}}]{ZiabariNC2018}%
  \BibitemOpen
  \bibfield  {author} {\bibinfo {author} {\bibfnamefont {A.}~\bibnamefont
  {Ziabari}}, \bibinfo {author} {\bibfnamefont {P.}~\bibnamefont {Torres}},
  \bibinfo {author} {\bibfnamefont {B.}~\bibnamefont {Vermeersch}}, \bibinfo
  {author} {\bibfnamefont {Y.}~\bibnamefont {Xuan}}, \bibinfo {author}
  {\bibfnamefont {X.}~\bibnamefont {Cartoix{\`a}}}, \bibinfo {author}
  {\bibfnamefont {A.}~\bibnamefont {Torell{\'o}}}, \bibinfo {author}
  {\bibfnamefont {J.-H.}\ \bibnamefont {Bahk}}, \bibinfo {author}
  {\bibfnamefont {Y.~R.}\ \bibnamefont {Koh}}, \bibinfo {author} {\bibfnamefont
  {M.}~\bibnamefont {Parsa}}, \bibinfo {author} {\bibfnamefont {P.~D.}\
  \bibnamefont {Ye}}, \bibinfo {author} {\bibfnamefont {F.~X.}\ \bibnamefont
  {Alvarez}}, \ and\ \bibinfo {author} {\bibfnamefont {A.}~\bibnamefont
  {Shakouri}},\ }\href {\doibase 10.1038/s41467-017-02652-4} {\bibfield
  {journal} {\bibinfo  {journal} {Nat. Commun.}\ }\textbf {\bibinfo {volume}
  {9}},\ \bibinfo {pages} {255} (\bibinfo {year} {2018})}\BibitemShut {NoStop}%
\bibitem [{\citenamefont {Beardo}\ \emph {et~al.}(2020)\citenamefont {Beardo},
  \citenamefont {Hennessy}, \citenamefont {Sendra}, \citenamefont {Camacho},
  \citenamefont {Myers}, \citenamefont {Bafaluy},\ and\ \citenamefont
  {Alvarez}}]{BeardoPRB2020}%
  \BibitemOpen
  \bibfield  {author} {\bibinfo {author} {\bibfnamefont {A.}~\bibnamefont
  {Beardo}}, \bibinfo {author} {\bibfnamefont {M.~G.}\ \bibnamefont
  {Hennessy}}, \bibinfo {author} {\bibfnamefont {L.}~\bibnamefont {Sendra}},
  \bibinfo {author} {\bibfnamefont {J.}~\bibnamefont {Camacho}}, \bibinfo
  {author} {\bibfnamefont {T.~G.}\ \bibnamefont {Myers}}, \bibinfo {author}
  {\bibfnamefont {J.}~\bibnamefont {Bafaluy}}, \ and\ \bibinfo {author}
  {\bibfnamefont {F.~X.}\ \bibnamefont {Alvarez}},\ }\href {\doibase
  10.1103/PhysRevB.101.075303} {\bibfield  {journal} {\bibinfo  {journal}
  {Phys. Rev. B}\ }\textbf {\bibinfo {volume} {101}},\ \bibinfo {pages}
  {075303} (\bibinfo {year} {2020})}\BibitemShut {NoStop}%
\bibitem [{\citenamefont {Johnson}\ \emph {et~al.}(2013)\citenamefont
  {Johnson}, \citenamefont {Maznev}, \citenamefont {Cuffe}, \citenamefont
  {Eliason}, \citenamefont {Minnich}, \citenamefont {Kehoe}, \citenamefont
  {{\relax C. M. Sotomayor Torres}}, \citenamefont {Chen},\ and\ \citenamefont
  {Nelson}}]{JohnsonPRL2013}%
  \BibitemOpen
  \bibfield  {author} {\bibinfo {author} {\bibfnamefont {J.~A.}\ \bibnamefont
  {Johnson}}, \bibinfo {author} {\bibfnamefont {A.~A.}\ \bibnamefont {Maznev}},
  \bibinfo {author} {\bibfnamefont {J.}~\bibnamefont {Cuffe}}, \bibinfo
  {author} {\bibfnamefont {J.~K.}\ \bibnamefont {Eliason}}, \bibinfo {author}
  {\bibfnamefont {A.~J.}\ \bibnamefont {Minnich}}, \bibinfo {author}
  {\bibfnamefont {T.}~\bibnamefont {Kehoe}}, \bibinfo {author} {\bibnamefont
  {{\relax C. M. Sotomayor Torres}}}, \bibinfo {author} {\bibfnamefont
  {G.}~\bibnamefont {Chen}}, \ and\ \bibinfo {author} {\bibfnamefont {K.~A.}\
  \bibnamefont {Nelson}},\ }\href {\doibase 10.1103/PhysRevLett.110.025901}
  {\bibfield  {journal} {\bibinfo  {journal} {Phys. Rev. Lett.}\ }\textbf
  {\bibinfo {volume} {110}},\ \bibinfo {pages} {025901} (\bibinfo {year}
  {2013})}\BibitemShut {NoStop}%
\bibitem [{\citenamefont {Torres}\ \emph {et~al.}(2018)\citenamefont {Torres},
  \citenamefont {Ziabari}, \citenamefont {Torell\'o}, \citenamefont {Bafaluy},
  \citenamefont {Camacho}, \citenamefont {Cartoix\`a}, \citenamefont
  {Shakouri},\ and\ \citenamefont {Alvarez}}]{TorresPRM2018}%
  \BibitemOpen
  \bibfield  {author} {\bibinfo {author} {\bibfnamefont {P.}~\bibnamefont
  {Torres}}, \bibinfo {author} {\bibfnamefont {A.}~\bibnamefont {Ziabari}},
  \bibinfo {author} {\bibfnamefont {A.}~\bibnamefont {Torell\'o}}, \bibinfo
  {author} {\bibfnamefont {J.}~\bibnamefont {Bafaluy}}, \bibinfo {author}
  {\bibfnamefont {J.}~\bibnamefont {Camacho}}, \bibinfo {author} {\bibfnamefont
  {X.}~\bibnamefont {Cartoix\`a}}, \bibinfo {author} {\bibfnamefont
  {A.}~\bibnamefont {Shakouri}}, \ and\ \bibinfo {author} {\bibfnamefont
  {F.~X.}\ \bibnamefont {Alvarez}},\ }\href {\doibase
  10.1103/PhysRevMaterials.2.076001} {\bibfield  {journal} {\bibinfo  {journal}
  {Phys. Rev. Materials}\ }\textbf {\bibinfo {volume} {2}},\ \bibinfo {pages}
  {076001} (\bibinfo {year} {2018})}\BibitemShut {NoStop}%
\bibitem [{\citenamefont {Joseph}\ and\ \citenamefont
  {Preziosi}(1989)}]{JosephRMP1989}%
  \BibitemOpen
  \bibfield  {author} {\bibinfo {author} {\bibfnamefont {D.~D.}\ \bibnamefont
  {Joseph}}\ and\ \bibinfo {author} {\bibfnamefont {L.}~\bibnamefont
  {Preziosi}},\ }\href {\doibase 10.1103/RevModPhys.61.41} {\bibfield
  {journal} {\bibinfo  {journal} {Rev. Mod. Phys.}\ }\textbf {\bibinfo {volume}
  {61}},\ \bibinfo {pages} {41} (\bibinfo {year} {1989})}\BibitemShut {NoStop}%
\bibitem [{\citenamefont {Christov}\ and\ \citenamefont
  {Jordan}(2005)}]{ChristovPRL2005}%
  \BibitemOpen
  \bibfield  {author} {\bibinfo {author} {\bibfnamefont {C.~I.}\ \bibnamefont
  {Christov}}\ and\ \bibinfo {author} {\bibfnamefont {P.~M.}\ \bibnamefont
  {Jordan}},\ }\href {\doibase 10.1103/PhysRevLett.94.154301} {\bibfield
  {journal} {\bibinfo  {journal} {Phys. Rev. Lett.}\ }\textbf {\bibinfo
  {volume} {94}},\ \bibinfo {pages} {154301} (\bibinfo {year}
  {2005})}\BibitemShut {NoStop}%
\bibitem [{\citenamefont {Torres~Alvarez}(2018)}]{PolThesis}%
  \BibitemOpen
  \bibfield  {author} {\bibinfo {author} {\bibfnamefont {P.}~\bibnamefont
  {Torres~Alvarez}},\ }\emph {\bibinfo {title} {Thermal Transport in
  Semiconductors: First Principles and Phonon Hydrodynamics}},\ \href@noop {}
  {Ph.D. thesis},\ \bibinfo  {school} {Universitat Aut\`onoma de Barcelona},
  \bibinfo {address} {Bellaterra, Spain} (\bibinfo {year} {2018})\BibitemShut
  {NoStop}%
\bibitem [{\citenamefont {Guyer}\ and\ \citenamefont
  {Krumhansl}(1966)}]{GuyerPR1966}%
  \BibitemOpen
  \bibfield  {author} {\bibinfo {author} {\bibfnamefont {R.~A.}\ \bibnamefont
  {Guyer}}\ and\ \bibinfo {author} {\bibfnamefont {J.~A.}\ \bibnamefont
  {Krumhansl}},\ }\href {\doibase 10.1103/PhysRev.148.766} {\bibfield
  {journal} {\bibinfo  {journal} {Phys. Rev.}\ }\textbf {\bibinfo {volume}
  {148}},\ \bibinfo {pages} {766} (\bibinfo {year} {1966})}\BibitemShut
  {NoStop}%
\bibitem [{\citenamefont {Guo}\ and\ \citenamefont {Wang}(2015)}]{GuoPR2015}%
  \BibitemOpen
  \bibfield  {author} {\bibinfo {author} {\bibfnamefont {Y.}~\bibnamefont
  {Guo}}\ and\ \bibinfo {author} {\bibfnamefont {M.}~\bibnamefont {Wang}},\
  }\href {\doibase https://doi.org/10.1016/j.physrep.2015.07.003} {\bibfield
  {journal} {\bibinfo  {journal} {Phys. Rep.}\ }\textbf {\bibinfo {volume}
  {595}},\ \bibinfo {pages} {1} (\bibinfo {year} {2015})}\BibitemShut {NoStop}%
\bibitem [{\citenamefont {Ziman}(2001)}]{ZimanEPH}%
  \BibitemOpen
  \bibfield  {author} {\bibinfo {author} {\bibfnamefont {J.~M.}\ \bibnamefont
  {Ziman}},\ }\href {\doibase 10.1093/acprof:oso/9780198507796.001.0001} {\emph
  {\bibinfo {title} {Electrons and Phonons: The Theory of Transport Phenomena
  in Solids}}},\ Oxford Classic Texts in the Physical Sciences\ (\bibinfo
  {publisher} {Oxford University Press},\ \bibinfo {address} {Oxford},\
  \bibinfo {year} {2001})\ p.\ \bibinfo {pages} {568}\BibitemShut {NoStop}%
\bibitem [{\citenamefont {Li}\ \emph {et~al.}(2012)\citenamefont {Li},
  \citenamefont {Mingo}, \citenamefont {Lindsay}, \citenamefont {Broido},
  \citenamefont {Stewart},\ and\ \citenamefont {Katcho}}]{LiPRB2012}%
  \BibitemOpen
  \bibfield  {author} {\bibinfo {author} {\bibfnamefont {W.}~\bibnamefont
  {Li}}, \bibinfo {author} {\bibfnamefont {N.}~\bibnamefont {Mingo}}, \bibinfo
  {author} {\bibfnamefont {L.}~\bibnamefont {Lindsay}}, \bibinfo {author}
  {\bibfnamefont {D.~A.}\ \bibnamefont {Broido}}, \bibinfo {author}
  {\bibfnamefont {D.~A.}\ \bibnamefont {Stewart}}, \ and\ \bibinfo {author}
  {\bibfnamefont {N.~A.}\ \bibnamefont {Katcho}},\ }\href {\doibase
  10.1103/PhysRevB.85.195436} {\bibfield  {journal} {\bibinfo  {journal} {Phys.
  Rev. B}\ }\textbf {\bibinfo {volume} {85}},\ \bibinfo {pages} {195436}
  (\bibinfo {year} {2012})}\BibitemShut {NoStop}%
\bibitem [{\citenamefont {Guo}\ and\ \citenamefont {Wang}(2018)}]{GuoPRB2018}%
  \BibitemOpen
  \bibfield  {author} {\bibinfo {author} {\bibfnamefont {Y.}~\bibnamefont
  {Guo}}\ and\ \bibinfo {author} {\bibfnamefont {M.}~\bibnamefont {Wang}},\
  }\href {\doibase 10.1103/PhysRevB.97.035421} {\bibfield  {journal} {\bibinfo
  {journal} {Phys. Rev. B}\ }\textbf {\bibinfo {volume} {97}},\ \bibinfo
  {pages} {035421} (\bibinfo {year} {2018})}\BibitemShut {NoStop}%
\bibitem [{\citenamefont {Sendra}\ \emph {et~al.}(2021)\citenamefont {Sendra},
  \citenamefont {Beardo}, \citenamefont {Torres}, \citenamefont {Bafaluy},
  \citenamefont {Alvarez},\ and\ \citenamefont {Camacho}}]{SendraPRB2021}%
  \BibitemOpen
  \bibfield  {author} {\bibinfo {author} {\bibfnamefont {L.}~\bibnamefont
  {Sendra}}, \bibinfo {author} {\bibfnamefont {A.}~\bibnamefont {Beardo}},
  \bibinfo {author} {\bibfnamefont {P.}~\bibnamefont {Torres}}, \bibinfo
  {author} {\bibfnamefont {J.}~\bibnamefont {Bafaluy}}, \bibinfo {author}
  {\bibfnamefont {F.~X.}\ \bibnamefont {Alvarez}}, \ and\ \bibinfo {author}
  {\bibfnamefont {J.}~\bibnamefont {Camacho}},\ }\href {\doibase
  10.1103/PhysRevB.103.L140301} {\bibfield  {journal} {\bibinfo  {journal}
  {Phys. Rev. B}\ }\textbf {\bibinfo {volume} {103}},\ \bibinfo {pages}
  {L140301} (\bibinfo {year} {2021})}\BibitemShut {NoStop}%
\bibitem [{\citenamefont {Levitov}\ and\ \citenamefont
  {Falkovich}(2016)}]{Levitov2016NP}%
  \BibitemOpen
  \bibfield  {author} {\bibinfo {author} {\bibfnamefont {L.}~\bibnamefont
  {Levitov}}\ and\ \bibinfo {author} {\bibfnamefont {G.}~\bibnamefont
  {Falkovich}},\ }\href {\doibase 10.1038/nphys3667} {\bibfield  {journal}
  {\bibinfo  {journal} {Nat. Phys.}\ }\textbf {\bibinfo {volume} {12}},\
  \bibinfo {pages} {672} (\bibinfo {year} {2016})}\BibitemShut {NoStop}%
\bibitem [{\citenamefont {Zhang}\ \emph {et~al.}(2019)\citenamefont {Zhang},
  \citenamefont {Guo},\ and\ \citenamefont {Chen}}]{ZhangIJHMT2019}%
  \BibitemOpen
  \bibfield  {author} {\bibinfo {author} {\bibfnamefont {C.}~\bibnamefont
  {Zhang}}, \bibinfo {author} {\bibfnamefont {Z.}~\bibnamefont {Guo}}, \ and\
  \bibinfo {author} {\bibfnamefont {S.}~\bibnamefont {Chen}},\ }\href {\doibase
  https://doi.org/10.1016/j.ijheatmasstransfer.2018.10.141} {\bibfield
  {journal} {\bibinfo  {journal} {Int. J. Heat Mass Transf.}\ }\textbf
  {\bibinfo {volume} {130}},\ \bibinfo {pages} {1366} (\bibinfo {year}
  {2019})}\BibitemShut {NoStop}%
\bibitem [{\citenamefont {Shang}\ \emph {et~al.}(2020)\citenamefont {Shang},
  \citenamefont {Zhang}, \citenamefont {Guo},\ and\ \citenamefont
  {L{\"u}}}]{Shang2020SR}%
  \BibitemOpen
  \bibfield  {author} {\bibinfo {author} {\bibfnamefont {M.-Y.}\ \bibnamefont
  {Shang}}, \bibinfo {author} {\bibfnamefont {C.}~\bibnamefont {Zhang}},
  \bibinfo {author} {\bibfnamefont {Z.}~\bibnamefont {Guo}}, \ and\ \bibinfo
  {author} {\bibfnamefont {J.-T.}\ \bibnamefont {L{\"u}}},\ }\href {\doibase
  10.1038/s41598-020-65221-8} {\bibfield  {journal} {\bibinfo  {journal} {Sci.
  Rep.}\ }\textbf {\bibinfo {volume} {10}},\ \bibinfo {pages} {8272} (\bibinfo
  {year} {2020})}\BibitemShut {NoStop}%
\bibitem [{\citenamefont {Zhang}\ \emph
  {et~al.}(2021{\natexlab{a}})\citenamefont {Zhang}, \citenamefont {Chen},\
  and\ \citenamefont {Guo}}]{ZhangIJHMT2021}%
  \BibitemOpen
  \bibfield  {author} {\bibinfo {author} {\bibfnamefont {C.}~\bibnamefont
  {Zhang}}, \bibinfo {author} {\bibfnamefont {S.}~\bibnamefont {Chen}}, \ and\
  \bibinfo {author} {\bibfnamefont {Z.}~\bibnamefont {Guo}},\ }\href {\doibase
  https://doi.org/10.1016/j.ijheatmasstransfer.2021.121282} {\bibfield
  {journal} {\bibinfo  {journal} {Int. J. Heat Mass Transf.}\ }\textbf
  {\bibinfo {volume} {176}},\ \bibinfo {pages} {121282} (\bibinfo {year}
  {2021}{\natexlab{a}})}\BibitemShut {NoStop}%
\bibitem [{\citenamefont {Guo}\ \emph {et~al.}(2021{\natexlab{a}})\citenamefont
  {Guo}, \citenamefont {Zhang}, \citenamefont {Nomura}, \citenamefont {Volz},\
  and\ \citenamefont {Wang}}]{GuoIJHMT2021}%
  \BibitemOpen
  \bibfield  {author} {\bibinfo {author} {\bibfnamefont {Y.}~\bibnamefont
  {Guo}}, \bibinfo {author} {\bibfnamefont {Z.}~\bibnamefont {Zhang}}, \bibinfo
  {author} {\bibfnamefont {M.}~\bibnamefont {Nomura}}, \bibinfo {author}
  {\bibfnamefont {S.}~\bibnamefont {Volz}}, \ and\ \bibinfo {author}
  {\bibfnamefont {M.}~\bibnamefont {Wang}},\ }\href
  {https://www.sciencedirect.com/science/article/pii/S0017931021000843}
  {\bibfield  {journal} {\bibinfo  {journal} {Int. J. Heat Mass Transf.}\
  }\textbf {\bibinfo {volume} {169}},\ \bibinfo {pages} {120981} (\bibinfo
  {year} {2021}{\natexlab{a}})}\BibitemShut {NoStop}%
\bibitem [{\citenamefont {Callaway}(1959)}]{CallawayPR1959}%
  \BibitemOpen
  \bibfield  {author} {\bibinfo {author} {\bibfnamefont {J.}~\bibnamefont
  {Callaway}},\ }\href {\doibase 10.1103/PhysRev.113.1046} {\bibfield
  {journal} {\bibinfo  {journal} {Phys. Rev.}\ }\textbf {\bibinfo {volume}
  {113}},\ \bibinfo {pages} {1046} (\bibinfo {year} {1959})}\BibitemShut
  {NoStop}%
\bibitem [{\citenamefont {Ding}\ \emph {et~al.}(2018)\citenamefont {Ding},
  \citenamefont {Zhou}, \citenamefont {Song}, \citenamefont {Li}, \citenamefont
  {Liu},\ and\ \citenamefont {Chen}}]{DingPRB2018}%
  \BibitemOpen
  \bibfield  {author} {\bibinfo {author} {\bibfnamefont {Z.}~\bibnamefont
  {Ding}}, \bibinfo {author} {\bibfnamefont {J.}~\bibnamefont {Zhou}}, \bibinfo
  {author} {\bibfnamefont {B.}~\bibnamefont {Song}}, \bibinfo {author}
  {\bibfnamefont {M.}~\bibnamefont {Li}}, \bibinfo {author} {\bibfnamefont
  {T.-H.}\ \bibnamefont {Liu}}, \ and\ \bibinfo {author} {\bibfnamefont
  {G.}~\bibnamefont {Chen}},\ }\href {\doibase 10.1103/PhysRevB.98.180302}
  {\bibfield  {journal} {\bibinfo  {journal} {Phys. Rev. B}\ }\textbf {\bibinfo
  {volume} {98}},\ \bibinfo {pages} {180302(R)} (\bibinfo {year}
  {2018})}\BibitemShut {NoStop}%
\bibitem [{\citenamefont {Guo}\ \emph {et~al.}(2021{\natexlab{b}})\citenamefont
  {Guo}, \citenamefont {Zhang}, \citenamefont {Bescond}, \citenamefont {Xiong},
  \citenamefont {Wang}, \citenamefont {Nomura},\ and\ \citenamefont
  {Volz}}]{GuoPRB2021}%
  \BibitemOpen
  \bibfield  {author} {\bibinfo {author} {\bibfnamefont {Y.}~\bibnamefont
  {Guo}}, \bibinfo {author} {\bibfnamefont {Z.}~\bibnamefont {Zhang}}, \bibinfo
  {author} {\bibfnamefont {M.}~\bibnamefont {Bescond}}, \bibinfo {author}
  {\bibfnamefont {S.}~\bibnamefont {Xiong}}, \bibinfo {author} {\bibfnamefont
  {M.}~\bibnamefont {Wang}}, \bibinfo {author} {\bibfnamefont {M.}~\bibnamefont
  {Nomura}}, \ and\ \bibinfo {author} {\bibfnamefont {S.}~\bibnamefont
  {Volz}},\ }\href {\doibase 10.1103/PhysRevB.104.075450} {\bibfield  {journal}
  {\bibinfo  {journal} {Phys. Rev. B}\ }\textbf {\bibinfo {volume} {104}},\
  \bibinfo {pages} {075450} (\bibinfo {year} {2021}{\natexlab{b}})}\BibitemShut
  {NoStop}%
\bibitem [{\citenamefont {Taylor}\ and\ \citenamefont
  {Heinonen}(2002)}]{taylor_heinonen_2002BOOK}%
  \BibitemOpen
  \bibfield  {author} {\bibinfo {author} {\bibfnamefont {P.~L.}\ \bibnamefont
  {Taylor}}\ and\ \bibinfo {author} {\bibfnamefont {O.}~\bibnamefont
  {Heinonen}},\ }\href {\doibase 10.1017/CBO9780511998782} {\emph {\bibinfo
  {title} {A Quantum Approach to Condensed Matter Physics}}}\ (\bibinfo
  {publisher} {Cambridge University Press},\ \bibinfo {year}
  {2002})\BibitemShut {NoStop}%
\bibitem [{\citenamefont {Maznev}\ and\ \citenamefont
  {Wright}(2014)}]{MaznevAJP2014}%
  \BibitemOpen
  \bibfield  {author} {\bibinfo {author} {\bibfnamefont {A.~A.}\ \bibnamefont
  {Maznev}}\ and\ \bibinfo {author} {\bibfnamefont {O.~B.}\ \bibnamefont
  {Wright}},\ }\href {\doibase 10.1119/1.4892612} {\bibfield  {journal}
  {\bibinfo  {journal} {Am. J. Phys.}\ }\textbf {\bibinfo {volume} {82}},\
  \bibinfo {pages} {1062} (\bibinfo {year} {2014})}\BibitemShut {NoStop}%
\bibitem [{\citenamefont {Landon}\ and\ \citenamefont
  {Hadjiconstantinou}(2014)}]{LandonJAP2014}%
  \BibitemOpen
  \bibfield  {author} {\bibinfo {author} {\bibfnamefont {C.}~\bibnamefont
  {Landon}}\ and\ \bibinfo {author} {\bibfnamefont {N.}~\bibnamefont
  {Hadjiconstantinou}},\ }\href@noop {} {\bibfield  {journal} {\bibinfo
  {journal} {J. Appl. Phys.}\ }\textbf {\bibinfo {volume} {116}},\ \bibinfo
  {pages} {163502} (\bibinfo {year} {2014})}\BibitemShut {NoStop}%
\bibitem [{\citenamefont {Li}\ and\ \citenamefont {Lee}(2018)}]{LiPRB2018}%
  \BibitemOpen
  \bibfield  {author} {\bibinfo {author} {\bibfnamefont {X.}~\bibnamefont
  {Li}}\ and\ \bibinfo {author} {\bibfnamefont {S.}~\bibnamefont {Lee}},\
  }\href {\doibase 10.1103/PhysRevB.97.094309} {\bibfield  {journal} {\bibinfo
  {journal} {Phys. Rev. B}\ }\textbf {\bibinfo {volume} {97}},\ \bibinfo
  {pages} {094309} (\bibinfo {year} {2018})}\BibitemShut {NoStop}%
\bibitem [{\citenamefont {Li}\ and\ \citenamefont {Lee}(2019)}]{LiPRB2019}%
  \BibitemOpen
  \bibfield  {author} {\bibinfo {author} {\bibfnamefont {X.}~\bibnamefont
  {Li}}\ and\ \bibinfo {author} {\bibfnamefont {S.}~\bibnamefont {Lee}},\
  }\href {\doibase 10.1103/PhysRevB.99.085202} {\bibfield  {journal} {\bibinfo
  {journal} {Phys. Rev. B}\ }\textbf {\bibinfo {volume} {99}},\ \bibinfo
  {pages} {085202} (\bibinfo {year} {2019})}\BibitemShut {NoStop}%
\bibitem [{\citenamefont {Cepellotti}\ \emph {et~al.}(2015)\citenamefont
  {Cepellotti}, \citenamefont {Fugallo}, \citenamefont {Paulatto},
  \citenamefont {Lazzeri}, \citenamefont {Mauri},\ and\ \citenamefont
  {Marzari}}]{CepellottiNatCom2015}%
  \BibitemOpen
  \bibfield  {author} {\bibinfo {author} {\bibfnamefont {A.}~\bibnamefont
  {Cepellotti}}, \bibinfo {author} {\bibfnamefont {G.}~\bibnamefont {Fugallo}},
  \bibinfo {author} {\bibfnamefont {L.}~\bibnamefont {Paulatto}}, \bibinfo
  {author} {\bibfnamefont {M.}~\bibnamefont {Lazzeri}}, \bibinfo {author}
  {\bibfnamefont {F.}~\bibnamefont {Mauri}}, \ and\ \bibinfo {author}
  {\bibfnamefont {N.}~\bibnamefont {Marzari}},\ }\href {\doibase
  10.1038/ncomms7400} {\bibfield  {journal} {\bibinfo  {journal} {Nat.
  Commun.}\ }\textbf {\bibinfo {volume} {6}},\ \bibinfo {pages} {6400}
  (\bibinfo {year} {2015})}\BibitemShut {NoStop}%
\bibitem [{\citenamefont {Chen}(2021)}]{ChenNatRP2021}%
  \BibitemOpen
  \bibfield  {author} {\bibinfo {author} {\bibfnamefont {G.}~\bibnamefont
  {Chen}},\ }\href {\doibase 10.1038/s42254-021-00334-1} {\bibfield  {journal}
  {\bibinfo  {journal} {Nat. Rev. Phys.}\ }\textbf {\bibinfo {volume} {3}},\
  \bibinfo {pages} {555} (\bibinfo {year} {2021})}\BibitemShut {NoStop}%
\bibitem [{\citenamefont {Lindsay}\ \emph {et~al.}(2019)\citenamefont
  {Lindsay}, \citenamefont {Katre}, \citenamefont {Cepellotti},\ and\
  \citenamefont {Mingo}}]{LindsayJAP2019}%
  \BibitemOpen
  \bibfield  {author} {\bibinfo {author} {\bibfnamefont {L.}~\bibnamefont
  {Lindsay}}, \bibinfo {author} {\bibfnamefont {A.}~\bibnamefont {Katre}},
  \bibinfo {author} {\bibfnamefont {A.}~\bibnamefont {Cepellotti}}, \ and\
  \bibinfo {author} {\bibfnamefont {N.}~\bibnamefont {Mingo}},\ }\href
  {\doibase 10.1063/1.5108651} {\bibfield  {journal} {\bibinfo  {journal} {J.
  Appl. Phys.}\ }\textbf {\bibinfo {volume} {126}},\ \bibinfo {pages} {050902}
  (\bibinfo {year} {2019})}\BibitemShut {NoStop}%
\bibitem [{\citenamefont {Haratipour}\ \emph {et~al.}(2016)\citenamefont
  {Haratipour}, \citenamefont {Namgung}, \citenamefont {Oh},\ and\
  \citenamefont {Koester}}]{HaratipourACSNano2016}%
  \BibitemOpen
  \bibfield  {author} {\bibinfo {author} {\bibfnamefont {N.}~\bibnamefont
  {Haratipour}}, \bibinfo {author} {\bibfnamefont {S.}~\bibnamefont {Namgung}},
  \bibinfo {author} {\bibfnamefont {S.-H.}\ \bibnamefont {Oh}}, \ and\ \bibinfo
  {author} {\bibfnamefont {S.~J.}\ \bibnamefont {Koester}},\ }\href {\doibase
  10.1021/acsnano.6b00482} {\bibfield  {journal} {\bibinfo  {journal} {ACS
  Nano}\ }\textbf {\bibinfo {volume} {10}},\ \bibinfo {pages} {3791} (\bibinfo
  {year} {2016})}\BibitemShut {NoStop}%
\bibitem [{\citenamefont {Ilatikhameneh}\ \emph {et~al.}(2016)\citenamefont
  {Ilatikhameneh}, \citenamefont {Ameen}, \citenamefont {Novakovic},
  \citenamefont {Tan}, \citenamefont {Klimeck},\ and\ \citenamefont
  {Rahman}}]{IlatikhamenehSR2016}%
  \BibitemOpen
  \bibfield  {author} {\bibinfo {author} {\bibfnamefont {H.}~\bibnamefont
  {Ilatikhameneh}}, \bibinfo {author} {\bibfnamefont {T.}~\bibnamefont
  {Ameen}}, \bibinfo {author} {\bibfnamefont {B.}~\bibnamefont {Novakovic}},
  \bibinfo {author} {\bibfnamefont {Y.}~\bibnamefont {Tan}}, \bibinfo {author}
  {\bibfnamefont {G.}~\bibnamefont {Klimeck}}, \ and\ \bibinfo {author}
  {\bibfnamefont {R.}~\bibnamefont {Rahman}},\ }\href {\doibase
  10.1038/srep31501} {\bibfield  {journal} {\bibinfo  {journal} {Sci. Rep.}\
  }\textbf {\bibinfo {volume} {6}},\ \bibinfo {pages} {31501} (\bibinfo {year}
  {2016})}\BibitemShut {NoStop}%
\bibitem [{\citenamefont {Li}\ \emph {et~al.}(2014)\citenamefont {Li},
  \citenamefont {Carrete}, \citenamefont {{A. Katcho}},\ and\ \citenamefont
  {Mingo}}]{ShengBTE}%
  \BibitemOpen
  \bibfield  {author} {\bibinfo {author} {\bibfnamefont {W.}~\bibnamefont
  {Li}}, \bibinfo {author} {\bibfnamefont {J.}~\bibnamefont {Carrete}},
  \bibinfo {author} {\bibfnamefont {N.}~\bibnamefont {{A. Katcho}}}, \ and\
  \bibinfo {author} {\bibfnamefont {N.}~\bibnamefont {Mingo}},\ }\href
  {\doibase https://doi.org/10.1016/j.cpc.2014.02.015} {\bibfield  {journal}
  {\bibinfo  {journal} {Comput. Phys. Commun.}\ }\textbf {\bibinfo {volume}
  {185}},\ \bibinfo {pages} {1747} (\bibinfo {year} {2014})}\BibitemShut
  {NoStop}%
\bibitem [{\citenamefont {P\'eraud}\ and\ \citenamefont
  {Hadjiconstantinou}(2011)}]{PeraudPRB2011}%
  \BibitemOpen
  \bibfield  {author} {\bibinfo {author} {\bibfnamefont {J.-P.~M.}\
  \bibnamefont {P\'eraud}}\ and\ \bibinfo {author} {\bibfnamefont {N.~G.}\
  \bibnamefont {Hadjiconstantinou}},\ }\href {\doibase
  10.1103/PhysRevB.84.205331} {\bibfield  {journal} {\bibinfo  {journal} {Phys.
  Rev. B}\ }\textbf {\bibinfo {volume} {84}},\ \bibinfo {pages} {205331}
  (\bibinfo {year} {2011})}\BibitemShut {NoStop}%
\bibitem [{\citenamefont {P\'eraud}\ and\ \citenamefont
  {Hadjiconstantinou}(2012)}]{PeraudAPL2012}%
  \BibitemOpen
  \bibfield  {author} {\bibinfo {author} {\bibfnamefont {J.-P.~M.}\
  \bibnamefont {P\'eraud}}\ and\ \bibinfo {author} {\bibfnamefont {N.~G.}\
  \bibnamefont {Hadjiconstantinou}},\ }\href {\doibase 10.1063/1.4757607}
  {\bibfield  {journal} {\bibinfo  {journal} {Appl. Phys. Lett.}\ }\textbf
  {\bibinfo {volume} {101}},\ \bibinfo {pages} {153114} (\bibinfo {year}
  {2012})}\BibitemShut {NoStop}%
\bibitem [{\citenamefont {P{\'e}raud}\ \emph {et~al.}(2014)\citenamefont
  {P{\'e}raud}, \citenamefont {Landon},\ and\ \citenamefont
  {Hadjiconstantinou}}]{PeraudARHT2014}%
  \BibitemOpen
  \bibfield  {author} {\bibinfo {author} {\bibfnamefont {J.-P.~M.}\
  \bibnamefont {P{\'e}raud}}, \bibinfo {author} {\bibfnamefont {C.~D.}\
  \bibnamefont {Landon}}, \ and\ \bibinfo {author} {\bibfnamefont {N.~G.}\
  \bibnamefont {Hadjiconstantinou}},\ }\href@noop {} {\bibfield  {journal}
  {\bibinfo  {journal} {Annu. Rev. Heat Transf.}\ }\textbf {\bibinfo {volume}
  {17}} (\bibinfo {year} {2014})}\BibitemShut {NoStop}%
\bibitem [{\citenamefont {Landon}(2014)}]{LandonThesis}%
  \BibitemOpen
  \bibfield  {author} {\bibinfo {author} {\bibfnamefont {C.~D.}\ \bibnamefont
  {Landon}},\ }\emph {\bibinfo {title} {A deviational Monte Carlo formulation
  of ab initio phonon transport and its application to the study of kinetic
  effects in graphene ribbons}},\ \href@noop {} {Ph.D. thesis},\ \bibinfo
  {school} {Massachusetts Institute of Technology}, \bibinfo {address}
  {Cambridge, MA} (\bibinfo {year} {2014})\BibitemShut {NoStop}%
\bibitem [{\citenamefont {Raya-Moreno}\ \emph {et~al.}(2022)\citenamefont
  {Raya-Moreno}, \citenamefont {Cartoix{\`a}},\ and\ \citenamefont
  {Carrete}}]{myself}%
  \BibitemOpen
  \bibfield  {author} {\bibinfo {author} {\bibfnamefont {M.}~\bibnamefont
  {Raya-Moreno}}, \bibinfo {author} {\bibfnamefont {X.}~\bibnamefont
  {Cartoix{\`a}}}, \ and\ \bibinfo {author} {\bibfnamefont {J.}~\bibnamefont
  {Carrete}},\ }\href@noop {} {\bibfield  {journal} {\bibinfo  {journal} {arXiv
  preprint arXiv:2202.00505v2}\ } (\bibinfo {year} {2022})}\BibitemShut
  {NoStop}%
\bibitem [{alm()}]{almadatabase}%
  \BibitemOpen
  \href@noop {} {\enquote {\bibinfo {title} {almabte database},}\ }\bibinfo
  {howpublished} {\url{https://almabte.bitbucket.io/database/}},\ \bibinfo
  {note} {accessed: 2021-09-30}\BibitemShut {NoStop}%
\bibitem [{\citenamefont {Smith}\ \emph {et~al.}(2017)\citenamefont {Smith},
  \citenamefont {Vermeersch}, \citenamefont {Carrete}, \citenamefont {Ou},
  \citenamefont {Kim}, \citenamefont {Mingo}, \citenamefont {Akinwande},\ and\
  \citenamefont {Shi}}]{SmithAM2017}%
  \BibitemOpen
  \bibfield  {author} {\bibinfo {author} {\bibfnamefont {B.}~\bibnamefont
  {Smith}}, \bibinfo {author} {\bibfnamefont {B.}~\bibnamefont {Vermeersch}},
  \bibinfo {author} {\bibfnamefont {J.}~\bibnamefont {Carrete}}, \bibinfo
  {author} {\bibfnamefont {E.}~\bibnamefont {Ou}}, \bibinfo {author}
  {\bibfnamefont {J.}~\bibnamefont {Kim}}, \bibinfo {author} {\bibfnamefont
  {N.}~\bibnamefont {Mingo}}, \bibinfo {author} {\bibfnamefont
  {D.}~\bibnamefont {Akinwande}}, \ and\ \bibinfo {author} {\bibfnamefont
  {L.}~\bibnamefont {Shi}},\ }\href {\doibase
  https://doi.org/10.1002/adma.201603756} {\bibfield  {journal} {\bibinfo
  {journal} {Adv. Mater.}\ }\textbf {\bibinfo {volume} {29}},\ \bibinfo {pages}
  {1603756} (\bibinfo {year} {2017})}\BibitemShut {NoStop}%
\bibitem [{\citenamefont {Carrete}\ \emph {et~al.}(2016)\citenamefont
  {Carrete}, \citenamefont {Li}, \citenamefont {Lindsay}, \citenamefont
  {Broido}, \citenamefont {Gallego},\ and\ \citenamefont
  {Mingo}}]{CarreteMRL2016}%
  \BibitemOpen
  \bibfield  {author} {\bibinfo {author} {\bibfnamefont {J.}~\bibnamefont
  {Carrete}}, \bibinfo {author} {\bibfnamefont {W.}~\bibnamefont {Li}},
  \bibinfo {author} {\bibfnamefont {L.}~\bibnamefont {Lindsay}}, \bibinfo
  {author} {\bibfnamefont {D.~A.}\ \bibnamefont {Broido}}, \bibinfo {author}
  {\bibfnamefont {L.~J.}\ \bibnamefont {Gallego}}, \ and\ \bibinfo {author}
  {\bibfnamefont {N.}~\bibnamefont {Mingo}},\ }\href {\doibase
  10.1080/21663831.2016.1174163} {\bibfield  {journal} {\bibinfo  {journal}
  {Mater. Res. Lett.}\ }\textbf {\bibinfo {volume} {4}},\ \bibinfo {pages}
  {204} (\bibinfo {year} {2016})}\BibitemShut {NoStop}%
\bibitem [{\citenamefont {Castellanos-Gomez}\ \emph {et~al.}(2014)\citenamefont
  {Castellanos-Gomez}, \citenamefont {Vicarelli}, \citenamefont {Prada},
  \citenamefont {Island}, \citenamefont {Narasimha-Acharya}, \citenamefont
  {Blanter}, \citenamefont {Groenendijk}, \citenamefont {Buscema},
  \citenamefont {Steele}, \citenamefont {Alvarez}, \citenamefont {Zandbergen},
  \citenamefont {Palacios},\ and\ \citenamefont {van~der
  Zant}}]{CastellanosGomez2DMat2014}%
  \BibitemOpen
  \bibfield  {author} {\bibinfo {author} {\bibfnamefont {A.}~\bibnamefont
  {Castellanos-Gomez}}, \bibinfo {author} {\bibfnamefont {L.}~\bibnamefont
  {Vicarelli}}, \bibinfo {author} {\bibfnamefont {E.}~\bibnamefont {Prada}},
  \bibinfo {author} {\bibfnamefont {J.~O.}\ \bibnamefont {Island}}, \bibinfo
  {author} {\bibfnamefont {K.~L.}\ \bibnamefont {Narasimha-Acharya}}, \bibinfo
  {author} {\bibfnamefont {S.~I.}\ \bibnamefont {Blanter}}, \bibinfo {author}
  {\bibfnamefont {D.~J.}\ \bibnamefont {Groenendijk}}, \bibinfo {author}
  {\bibfnamefont {M.}~\bibnamefont {Buscema}}, \bibinfo {author} {\bibfnamefont
  {G.~A.}\ \bibnamefont {Steele}}, \bibinfo {author} {\bibfnamefont {J.~V.}\
  \bibnamefont {Alvarez}}, \bibinfo {author} {\bibfnamefont {H.~W.}\
  \bibnamefont {Zandbergen}}, \bibinfo {author} {\bibfnamefont {J.~J.}\
  \bibnamefont {Palacios}}, \ and\ \bibinfo {author} {\bibfnamefont {H.~S.~J.}\
  \bibnamefont {van~der Zant}},\ }\href {\doibase 10.1088/2053-1583/1/2/025001}
  {\bibfield  {journal} {\bibinfo  {journal} {2D Mater.}\ }\textbf {\bibinfo
  {volume} {1}},\ \bibinfo {pages} {025001} (\bibinfo {year}
  {2014})}\BibitemShut {NoStop}%
\bibitem [{\citenamefont {Logg}\ \emph {et~al.}(2012)\citenamefont {Logg},
  \citenamefont {Mardal},\ and\ \citenamefont {Wells}}]{FeniCsBook}%
  \BibitemOpen
  \bibfield  {author} {\bibinfo {author} {\bibfnamefont {A.}~\bibnamefont
  {Logg}}, \bibinfo {author} {\bibfnamefont {K.-A.}\ \bibnamefont {Mardal}}, \
  and\ \bibinfo {author} {\bibfnamefont {G.}~\bibnamefont {Wells}},\
  }\href@noop {} {\emph {\bibinfo {title} {Automated solution of differential
  equations by the finite element method: The FEniCS book}}},\ Vol.~\bibinfo
  {volume} {84}\ (\bibinfo  {publisher} {Springer Science \& Business Media},\
  \bibinfo {year} {2012})\BibitemShut {NoStop}%
\bibitem [{\citenamefont {Aln{\ae}s}\ \emph {et~al.}(2015)\citenamefont
  {Aln{\ae}s}, \citenamefont {Blechta}, \citenamefont {Hake}, \citenamefont
  {Johansson}, \citenamefont {Kehlet}, \citenamefont {Logg}, \citenamefont
  {Richardson}, \citenamefont {Ring}, \citenamefont {Rognes},\ and\
  \citenamefont {Wells}}]{alnaes2015fenics}%
  \BibitemOpen
  \bibfield  {author} {\bibinfo {author} {\bibfnamefont {M.}~\bibnamefont
  {Aln{\ae}s}}, \bibinfo {author} {\bibfnamefont {J.}~\bibnamefont {Blechta}},
  \bibinfo {author} {\bibfnamefont {J.}~\bibnamefont {Hake}}, \bibinfo {author}
  {\bibfnamefont {A.}~\bibnamefont {Johansson}}, \bibinfo {author}
  {\bibfnamefont {B.}~\bibnamefont {Kehlet}}, \bibinfo {author} {\bibfnamefont
  {A.}~\bibnamefont {Logg}}, \bibinfo {author} {\bibfnamefont {C.}~\bibnamefont
  {Richardson}}, \bibinfo {author} {\bibfnamefont {J.}~\bibnamefont {Ring}},
  \bibinfo {author} {\bibfnamefont {M.~E.}\ \bibnamefont {Rognes}}, \ and\
  \bibinfo {author} {\bibfnamefont {G.~N.}\ \bibnamefont {Wells}},\ }\href@noop
  {} {\bibfield  {journal} {\bibinfo  {journal} {Arch. Num. Soft.}\ }\textbf
  {\bibinfo {volume} {3}} (\bibinfo {year} {2015})}\BibitemShut {NoStop}%
\bibitem [{SM()}]{SM}%
  \BibitemOpen
  \bibinfo {note} {See the Supplemental Material at [URL will be inserted by
  publisher] for the FEniCs python script (\texttt{fenics.py}), the videos of
  the time evolution of the temperature (\texttt{temperature\_short.mp4}) and
  heat flux (\texttt{flux\_short.mp4}) of Fig.~\ref{Fig:VortexFormation}, and
  the videos of the time evolution of the temperature
  (\texttt{temperature\_large.mp4}) and heat flux (\texttt{flux\_large.mp4}) of
  Fig.~\ref{Fig:VortexFormationLARGE}.}\BibitemShut {Stop}%
\bibitem [{\citenamefont {Balandin}\ \emph {et~al.}(2008)\citenamefont
  {Balandin}, \citenamefont {Ghosh}, \citenamefont {Bao}, \citenamefont
  {Calizo}, \citenamefont {Teweldebrhan}, \citenamefont {Miao},\ and\
  \citenamefont {Lau}}]{BalandinNL2008}%
  \BibitemOpen
  \bibfield  {author} {\bibinfo {author} {\bibfnamefont {A.~A.}\ \bibnamefont
  {Balandin}}, \bibinfo {author} {\bibfnamefont {S.}~\bibnamefont {Ghosh}},
  \bibinfo {author} {\bibfnamefont {W.}~\bibnamefont {Bao}}, \bibinfo {author}
  {\bibfnamefont {I.}~\bibnamefont {Calizo}}, \bibinfo {author} {\bibfnamefont
  {D.}~\bibnamefont {Teweldebrhan}}, \bibinfo {author} {\bibfnamefont
  {F.}~\bibnamefont {Miao}}, \ and\ \bibinfo {author} {\bibfnamefont {C.~N.}\
  \bibnamefont {Lau}},\ }\href@noop {} {\bibfield  {journal} {\bibinfo
  {journal} {Nano Lett.}\ }\textbf {\bibinfo {volume} {8}},\ \bibinfo {pages}
  {902} (\bibinfo {year} {2008})}\BibitemShut {NoStop}%
\bibitem [{\citenamefont {Ghosh}\ \emph {et~al.}(2008)\citenamefont {Ghosh},
  \citenamefont {Calizo}, \citenamefont {Teweldebrhan}, \citenamefont
  {Pokatilov}, \citenamefont {Nika}, \citenamefont {Balandin}, \citenamefont
  {Bao}, \citenamefont {Miao},\ and\ \citenamefont {Lau}}]{GhoshAPL2008}%
  \BibitemOpen
  \bibfield  {author} {\bibinfo {author} {\bibfnamefont {S.}~\bibnamefont
  {Ghosh}}, \bibinfo {author} {\bibfnamefont {I.}~\bibnamefont {Calizo}},
  \bibinfo {author} {\bibfnamefont {D.}~\bibnamefont {Teweldebrhan}}, \bibinfo
  {author} {\bibfnamefont {E.~P.}\ \bibnamefont {Pokatilov}}, \bibinfo {author}
  {\bibfnamefont {D.~L.}\ \bibnamefont {Nika}}, \bibinfo {author}
  {\bibfnamefont {A.~A.}\ \bibnamefont {Balandin}}, \bibinfo {author}
  {\bibfnamefont {W.}~\bibnamefont {Bao}}, \bibinfo {author} {\bibfnamefont
  {F.}~\bibnamefont {Miao}}, \ and\ \bibinfo {author} {\bibfnamefont {C.~N.}\
  \bibnamefont {Lau}},\ }\href {\doibase 10.1063/1.2907977} {\bibfield
  {journal} {\bibinfo  {journal} {Appl. Phys. Lett.}\ }\textbf {\bibinfo
  {volume} {92}},\ \bibinfo {pages} {151911} (\bibinfo {year}
  {2008})}\BibitemShut {NoStop}%
\bibitem [{\citenamefont {Zhu}\ \emph {et~al.}(2014)\citenamefont {Zhu},
  \citenamefont {Zhang},\ and\ \citenamefont {Li}}]{ZhuPRB2014}%
  \BibitemOpen
  \bibfield  {author} {\bibinfo {author} {\bibfnamefont {L.}~\bibnamefont
  {Zhu}}, \bibinfo {author} {\bibfnamefont {G.}~\bibnamefont {Zhang}}, \ and\
  \bibinfo {author} {\bibfnamefont {B.}~\bibnamefont {Li}},\ }\href {\doibase
  10.1103/PhysRevB.90.214302} {\bibfield  {journal} {\bibinfo  {journal} {Phys.
  Rev. B}\ }\textbf {\bibinfo {volume} {90}},\ \bibinfo {pages} {214302}
  (\bibinfo {year} {2014})}\BibitemShut {NoStop}%
\bibitem [{\citenamefont {Jain}\ and\ \citenamefont
  {McGaughey}(2015)}]{JainSR2015}%
  \BibitemOpen
  \bibfield  {author} {\bibinfo {author} {\bibfnamefont {A.}~\bibnamefont
  {Jain}}\ and\ \bibinfo {author} {\bibfnamefont {A.~J.~H.}\ \bibnamefont
  {McGaughey}},\ }\href {\doibase 10.1038/srep08501} {\bibfield  {journal}
  {\bibinfo  {journal} {Sci. Rep.}\ }\textbf {\bibinfo {volume} {5}},\ \bibinfo
  {pages} {8501} (\bibinfo {year} {2015})}\BibitemShut {NoStop}%
\bibitem [{Note1()}]{Note1}%
  \BibitemOpen
  \bibinfo {note} {In this and all the subsequent plots the color scale of the
  temperature profiles is sightly non-linear in the vicinity of 300 K to
  highlight the negative resistivity zones}\BibitemShut {NoStop}%
\bibitem [{\citenamefont {Mei}\ \emph {et~al.}(2014)\citenamefont {Mei},
  \citenamefont {Maurer}, \citenamefont {Aksamija},\ and\ \citenamefont
  {Knezevic}}]{MeiJAP2014}%
  \BibitemOpen
  \bibfield  {author} {\bibinfo {author} {\bibfnamefont {S.}~\bibnamefont
  {Mei}}, \bibinfo {author} {\bibfnamefont {L.~N.}\ \bibnamefont {Maurer}},
  \bibinfo {author} {\bibfnamefont {Z.}~\bibnamefont {Aksamija}}, \ and\
  \bibinfo {author} {\bibfnamefont {I.}~\bibnamefont {Knezevic}},\ }\href
  {\doibase 10.1063/1.4899235} {\bibfield  {journal} {\bibinfo  {journal} {J.
  Appl. Phys.}\ }\textbf {\bibinfo {volume} {116}},\ \bibinfo {pages} {164307}
  (\bibinfo {year} {2014})}\BibitemShut {NoStop}%
\bibitem [{\citenamefont {Zhang}\ \emph
  {et~al.}(2021{\natexlab{b}})\citenamefont {Zhang}, \citenamefont {Guo},
  \citenamefont {Bescond}, \citenamefont {Chen}, \citenamefont {Nomura},\ and\
  \citenamefont {Volz}}]{ZhangAPLMat2021}%
  \BibitemOpen
  \bibfield  {author} {\bibinfo {author} {\bibfnamefont {Z.}~\bibnamefont
  {Zhang}}, \bibinfo {author} {\bibfnamefont {Y.}~\bibnamefont {Guo}}, \bibinfo
  {author} {\bibfnamefont {M.}~\bibnamefont {Bescond}}, \bibinfo {author}
  {\bibfnamefont {J.}~\bibnamefont {Chen}}, \bibinfo {author} {\bibfnamefont
  {M.}~\bibnamefont {Nomura}}, \ and\ \bibinfo {author} {\bibfnamefont
  {S.}~\bibnamefont {Volz}},\ }\href {\doibase 10.1063/5.0059024} {\bibfield
  {journal} {\bibinfo  {journal} {APL Mater.}\ }\textbf {\bibinfo {volume}
  {9}},\ \bibinfo {pages} {081102} (\bibinfo {year}
  {2021}{\natexlab{b}})}\BibitemShut {NoStop}%
\end{thebibliography}%

\end{document}